\documentclass[aps,prl,twocolumn,groupedaddress,showpacs,floatfix,superscriptaddress,longbibliography]{revtex4-2}
\usepackage[plainpages=false,pdfpagelabels,colorlinks=true,linkcolor=red,urlcolor=blue,citecolor=blue,pdftitle={Title},pdfauthor={},pdfdisplaydoctitle=true,pdfduplex=DuplexFlipLongEdge]{hyperref}
\usepackage{siunitx}
\usepackage{mhchem}
\usepackage{epsfig}
\usepackage{graphicx}
\usepackage[utf8]{inputenc}
\usepackage{amsmath,amssymb}
\usepackage[dvipsnames]{xcolor}

\usepackage[normalem]{ulem}

\usepackage{tabularx}
\usepackage{amsmath}

\usepackage[T1]{fontenc}
\usepackage[utf8]{inputenc}  % 一般现代 LaTeX 已默认支持

\begin{document}

%% \title{Effect of Dilution on the Half-Filled Holstein Model\\
\title{Role of small-radius and high-electronegativity A-Site dopants in enhancing proton transport and stability of perovskite electrolytes}
\author{Hang Ma}
\affiliation{School of Physics and Astronomy, Beijing Normal University, and Key Laboratory of Multiscale Spin Physics (Beijing Normal University), Ministry of Education, Beijing 100875, China\\} 
\author{Ying Liang}
\email{liangying@hebtu.edu.cn}
\affiliation{College of Physics, Hebei Normal University, and Hebei Advanced Thin Films Laboratory, Shijiazhuang 050024, China}
\affiliation{School of Physics and Astronomy, Beijing Normal University, and Key Laboratory of Multiscale Spin Physics (Beijing Normal University), Ministry of Education, Beijing 100875, China\\}
\author{Tianxing Ma}
\email{txma@bnu.edu.cn}
\affiliation{School of Physics and Astronomy, Beijing Normal University, and Key Laboratory of Multiscale Spin Physics (Beijing Normal University), Ministry of Education, Beijing 100875, China\\}

\begin{abstract}
The practical application of BaCeO$_3$-based electrolytes is limited by their poor chemical stability in proton-conducting solid oxide fuel cells. Commonly employed B-site doping strategies typically improve proton transport with limited improvement in stability. Recent experiments show that A-site Ca doping can simultaneously enhance both properties. Here, through first-principles calculations and mechanistic analysis of Ca-doped BaCeO$_3$, we identify the synergistic roles of small-radius, high-electronegativity A-site dopants in governing proton transport and chemical stability in perovskite electrolytes. We show that the higher electronegativity of A-site dopant weakens the A-O ionic bonding, facilitating oxygen-vacancy formation and enhancing proton uptake by increasing the basicity. This weakened A-O interaction also suppresses the formation of impurity phases and reduces the adsorption strength of acidic gases such as CO$_2$ and SO$_2$. The lattice contraction induced by the smaller ionic radius improves thermal stability and can enhance proton diffusion in systems where proton transfer is the rate-limiting step. Furthermore, we find that Ca surface segregation can mitigate grain-boundary resistance effects. Our results demonstrate the advantages of A-site Ca doping in Ba-based electrolytes, clarify the mechanisms by which small-radius, high-electronegativity dopants influence proton transport and chemical stability, and provide guidance for the design of high-performance proton-conducting electrolytes.
\end{abstract}

\date{Version 16.0 -- \today}

\maketitle

%%%%%%%%%%%%%%%%%%%%%%%%%%%%%%%%%%%%%%%%%%%%%%%%%%%%%%%%%%%%%%%%%%%%%%%%
%%%%%%%%%%%%%%%%%%%%%%%%%%%%%%%%%%%%%%%%%%%%%%%%%%%%%%%%%%%%%%%%%%%%%%%%

\section{Introduction}
With the growing global demand for clean energy, hydrogen fuel cells, which use hydrogen with high energy density as a fuel, have attracted increasing attention\cite{abe2019hydrogen}. Among them, solid oxide fuel cells (SOFCs) have been extensively studied because of their high electrical efficiency (exceeding 50\%) and overall energy utilization efficiency of more than 80\%\cite{XU2022115175,BICER20203670}. Conventional oxide-ion-conducting SOFCs require high operating temperatures above 800~\si{\celsius}\cite{TIMURKUTLUK20161101,doi:10.1021/acsami.3c09025}, because their oxide-ion electrolyte exhibits a relatively high activation energy for ion diffusion. Such high temperatures lead to several issues, including high manufacturing costs, material degradation, and performance loss. In contrast, proton-conducting SOFCs (H-SOFCs) are more advantageous, as they exhibit high proton conductivity even at intermediate temperatures of 400-600~\si{\celsius}\cite{ding2020self}. This high ionic conductivity at lower operating temperatures mainly arises from the low activation energy for proton diffusion in the electrolyte materials. Therefore, selecting electrolyte materials with high proton conductivity is crucial for improving the electrochemical performance of proton-conducting SOFCs\cite{somekawa2016physicochemical}. BaCeO$_3$ is among the most extensively investigated perovskite-type electrolytes because of its relatively high ionic conductivity and good sinterability compared with other systems\cite{kreuer2003proton,medvedev2014baceo3}. However, it exhibits poor chemical stability in humid and acidic environments\cite{fang2008chemical,somekawa2017physicochemical}. In addition, despite its relatively low grain boundary resistance, its proton conductivity remains lower than the bulk conductivity of BaZrO$_3$\cite{kreuer2003proton,kreuer1999aspects}. 

Conventional B-site doping strategies (e.g., Y, Sm, Gd, and Nb) generally improve proton conductivity but provide limited enhancement in chemical stability\cite{amsif2011influence,zhang2011electrical,chen2009proton,he2021new, he2022surface}. And recent studies have shown that A-site doping can simultaneously improve the electrochemical properties of electrolyte and cathode materials\cite{dudek2016some,dudek2019ba0,luo2023chemical,kothandan2024effect,lu2018engineering,CAO2026153443}. In particular, A-site Ca doping has been demonstrated to enhance both proton conductivity and chemical stability in BaCeO$_3$-based systems. Dudek et al.\cite{dudek2016some} reported that partial substitution of Ba with Ca in (Ba$_{1-x}$Ca$_x$)(M$_{0.9}$Y$_{0.1}$)O$_3$(M = Ce,Zr) solid solutions improves the sinterability of the samples, and the highest ionic conductivity was achieved at doping levels of $x=0.02–0.05$. Furthermore, Dudek et al.\cite{dudek2019ba0} later demonstrated that Ba$_{0.95}$Ca$_{0.05}$Ce$_{0.9}$Y$_{0.1}$O$_{3}$ exhibits enhanced chemical stability in a CO$_2$ atmosphere. More recently, Luo et al. \cite{luo2023chemical} found that introducing a small amount of Ca$^{2+}$ into BaCe$_{0.8}$Gd$_{0.2}$O$_3$ enhances both proton conductivity and stability under humid conditions, with optimal performance achieved at a Ca concentration of 1\%. Consistently, D. Kothandan et al.\cite{kothandan2024effect} reported that Ca doping with a concentration of 0.05 in BaCe$_{0.8}$Nd$_{0.2}$O$_3$ leads to a bulk conductivity exceeding the grain boundary conductivity, thereby improving the overall conductivity. Moreover, previous studies  have demonstrated that small-radius, high-electronegativity dopants can enhance proton conductivity and chemical stability in proton-conducting oxides\cite{kreuer2001proton,kreuer2003proton,oikawa2015correlation,peng2025enhancement}. This suggests that the advantages of A-site Ca doping may originate from its smaller ionic radius and higher electronegativity compared with the host Ba ion. Accordingly, we perform a comprehensive first-principles investigation of the effects of A-site Ca doping on proton concentration, proton diffusion, and stability in BaCeO$_3$, aiming to elucidate the mechanisms underlying the effects of small-radius, high-electronegativity dopants.

\begin{figure}[htbp]
\centering
\includegraphics[scale=0.31]{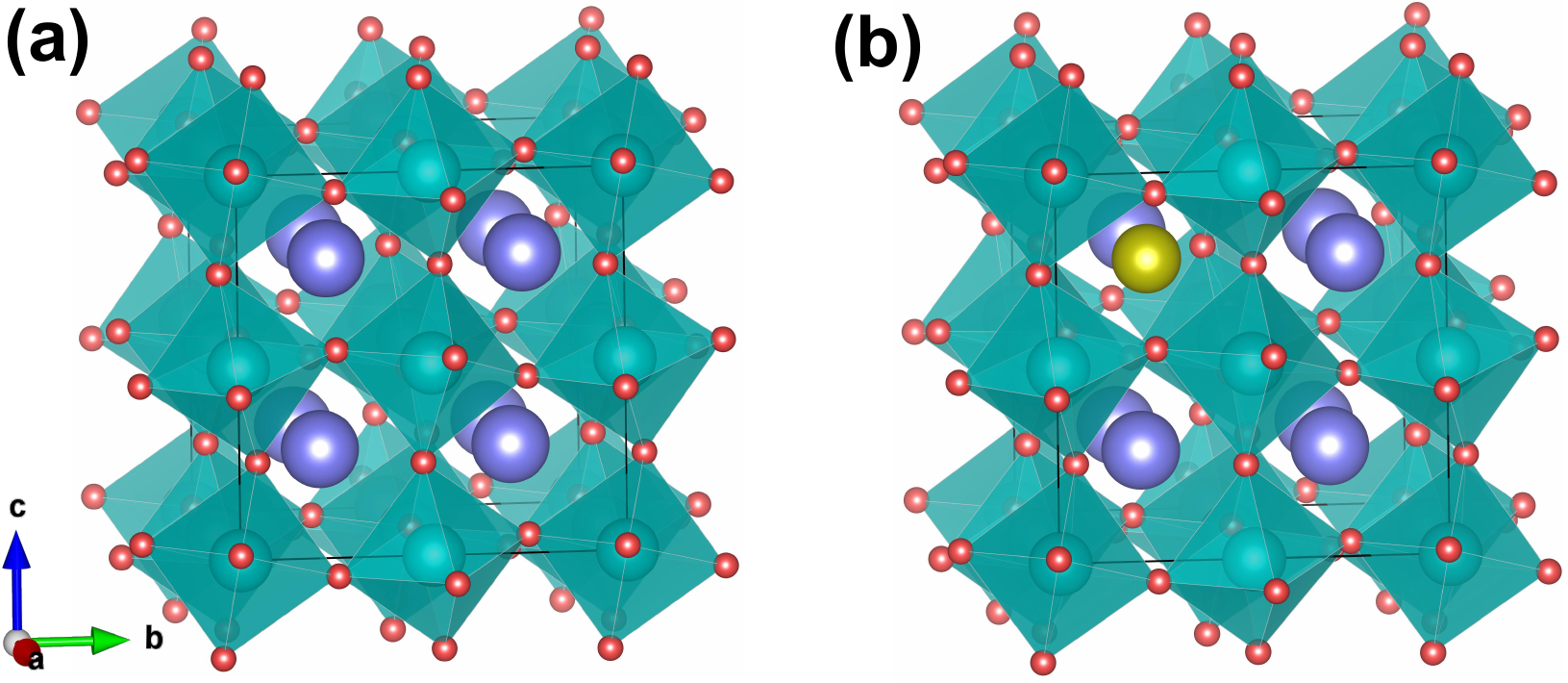}\\[0.5cm]
\includegraphics[scale=0.45]{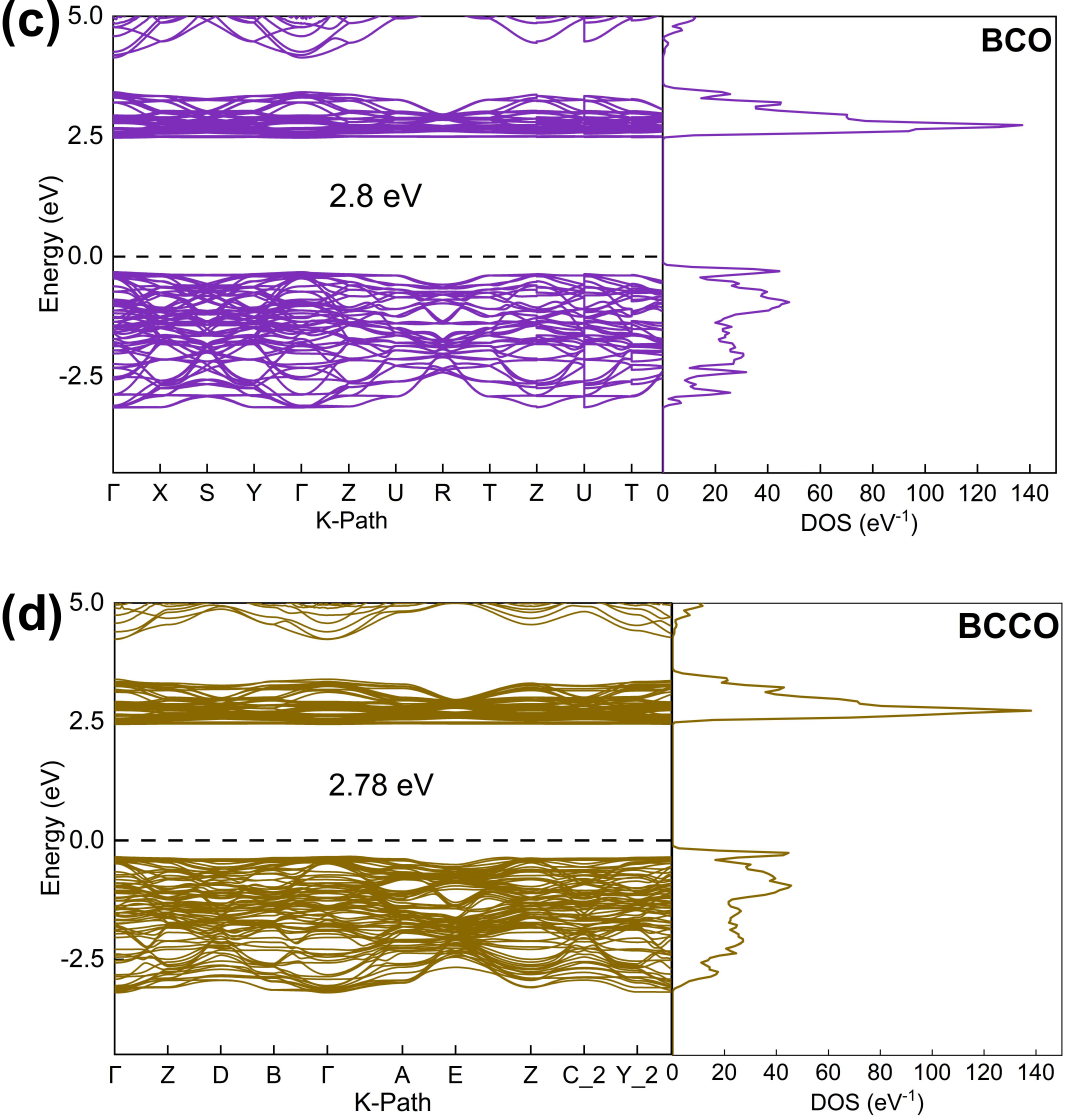}
\caption{(a) $\sqrt{2} \times \sqrt{2} \times \sqrt{2}$ supercell of BaCeO$_3$. (b) Supercell of Ba$_{0.875}$Ca$_{0.125}$CeO$_3$. The purple, yellow, cyan, and red spheres represent Ba, Ca, Ce, and O atoms, respectively. (c,d) Band structures and DOS of BCO and BCCO. }
\label{Fig1}
\end{figure}

From a theoretical perspective, B-site doping with transition metal elements can simultaneously alter the electronic structure and induce lattice distortions in the system\cite{zhang2018insight,gao2020designing}. The former mainly involves modifications of the electronic states near the Fermi level\cite{vignesh2024proton,liu2024yttrium}, where strong correlation and spin-orbit coupling (SOC) effects associated with the dopant atoms may lead to band splitting and the removal of degeneracy\cite{gracia2024quantum,he1997impedance,vignesh2025proton}, or introduce impurity levels within the band gap\cite{yoshino2000local,liu2020microscopic}. These effects can change the electronic density of states (DOS) at the Fermi level, influencing the degree of electron localization or delocalization and, consequently, the proton diffusion behavior. The latter effect arises from the size mismatch between the dopant and host B-site cations, leading to local lattice expansion or contraction\cite{PhysRevB.76.054307,bevillon2014dopant,rfs5-bnwt}. Such structural distortions can modify bond lengths and strengths, consequently altering the migration barriers of protons that couple to the lattice vibrations\cite{jing2020role,kreuer2000complexity,samgin2000lattice,du2020cooperative,rfs5-bnwt,n8dz-gwfj}. The two factors collectively affect proton diffusion. However, it is difficult to quantitatively distinguish their individual contributions, making it challenging to identify the dominant factor governing proton diffusion. In contrast, A-site Ca doping, as illustrated in Fig.~\ref{Fig1}(c,d), does not significantly modify the electronic structure, especially the density of states near the Fermi level, since Ca lacks 4f and 5d electrons. Therefore, the effect of lattice distortion can be isolated and independently examined with respect to proton diffusion.

In this work, we systematically investigate the effects of A-site Ca doping on the proton conductivity and chemical stability of BaCeO$_3$ using first-principles calculations. Our results successfully explain the experimentally observed improvements in both aspects. More importantly, we identify a dual advantage associated with small-radius, high-electronegativity dopants. The higher electronegativity weakens the A-O bond, facilitating oxygen-vacancy formation, and enhancing proton uptake by increasing the basicity, while simultaneously suppressing CO$_2$/SO$_2$ adsorption. The lattice contraction induced by the smaller ionic radius improves thermal stability and can enhance proton diffusion in systems where proton transfer is the rate-limiting step. These findings provide mechanistic insight and practical doping-design guidelines for optimizing proton-conducting electrolyte materials.

%%%%%%%%%%%%%%%%%%%%%%%%%%%%%%%%%%%%%%%%%%%%%%%%%%%%%%%%%%%%%%%%%%%%%%%%
%%%%%%%%%%%%%%%%%%%%%%%%%%%%%%%%%%%%%%%%%%%%%%%%%%%%%%%%%%%%%%%%%%%%%%%%

\section{Computational Method}
%%{\color{blue}
All DFT calculations were performed by the Vienna Ab Initio Simulation Package (VASP)\cite{kresse1996efficiency,PhysRevB.59.1758} with projector augmented wave (PAW) method\cite{kresse1996efficient}. The Perdew-Burke-Ernzerhof (PBE) based generalized gradient approximation (GGA)\cite{PhysRevLett.77.3865} exchange-correlation functional is used. The cutoff energy of the plane wave basis was set to 520~eV. And the electronic self-consistent field (SCF) convergence criterion was set to 10$^{-6}$~eV. Structural optimization was completed using the conjugate gradient algorithm until the Hellmann-Feynman forces on each atoms are less than 0.01~eV/\AA.

Based on the optimized orthorhombic BaCeO$_3$ unit cell, a $\sqrt{2} \times \sqrt{2} \times 1$ supercell containing 40 atoms (8 Ba, 8 Ce, and 24 O) was constructed to model the bulk BaCeO$_3$ (BCO) structure. Subsequently, one Ba atom was substituted by Ca to construct the bulk Ba$_{0.875}$Ca$_{0.125}$CeO$_3$(BCCO) structure, as illustrated in Fig.~\ref{Fig1}(a,b). All bulk supercell calculations employed a $3 \times 3 \times 3$ Monkhorst-Pack k-point mesh with a spacing of 0.035~\AA$^{-1}$. A seven-layer slab model containing 68 atoms was constructed to calculate the oxygen vacancy formation energy and adsorption energy. As shown in Fig.~\ref{Fig6}, a nonpolar BaO-terminated (001) surface, which is considered the most stable termination\cite{shishkin2012structural,heifets2007electronic,syha2012three}, was selected. A vacuum layer of 15~\AA\ was added to avoid interactions between adjacent surfaces\cite{tauer2013computational}. During the simulations, the bottom two layers were fixed, while the top five layers were fully relaxed. A $3 \times 3 \times 1$ Monkhorst-Pack k-point mesh was employed for Brillouin zone sampling. For all bulk and surface structure calculations, spin polarization was taken into account. To describe the on-site Coulomb interaction of the Ce 4f electrons, a Hubbard U value of 5~eV was applied to Ce\cite{liu2020microscopic,he2022surface}.

The migration energy barriers were calculated using the climbing-image nudged elastic band (CI-NEB) method\cite{henkelman2000climbing}. To better approximate realistic conditions, the initial and final states for proton rotation and transfer were fully relaxed. For the intermediate images, only the atomic positions were optimized while keeping the lattice parameters fixed, in order to avoid convergence issues during the calculations. The convergence criterion for the residual forces was set to 0.02 eV/\AA. It should be noted that the migration barriers reported in this work are obtained within the classical Born-Oppenheimer approximation. Nuclear quantum effects, which may further influence proton transfer, are not included here and could be addressed in future path-integral molecular dynamics studies. AIMD simulations were performed to investigate the thermal stability of the bulk BCO and BCCO systems at 800~K. 
The simulations were carried out for 10 ps with a time step of 0.5 fs. An NVT ensemble was employed using a Nos\'e-Hoover thermostat \cite{perdew1986density,yamaguchi2005thermoelectric} to maintain stable temperature control throughout the run. The $\Gamma$ point was used for Brillouin zone k-point sampling.

\section{Results and Discussion}
\subsection*{A. Crystal and Electron structure of BCO and BCCO}
\begin{figure}[htbp]
\raggedright
\includegraphics[width=0.4\textwidth]{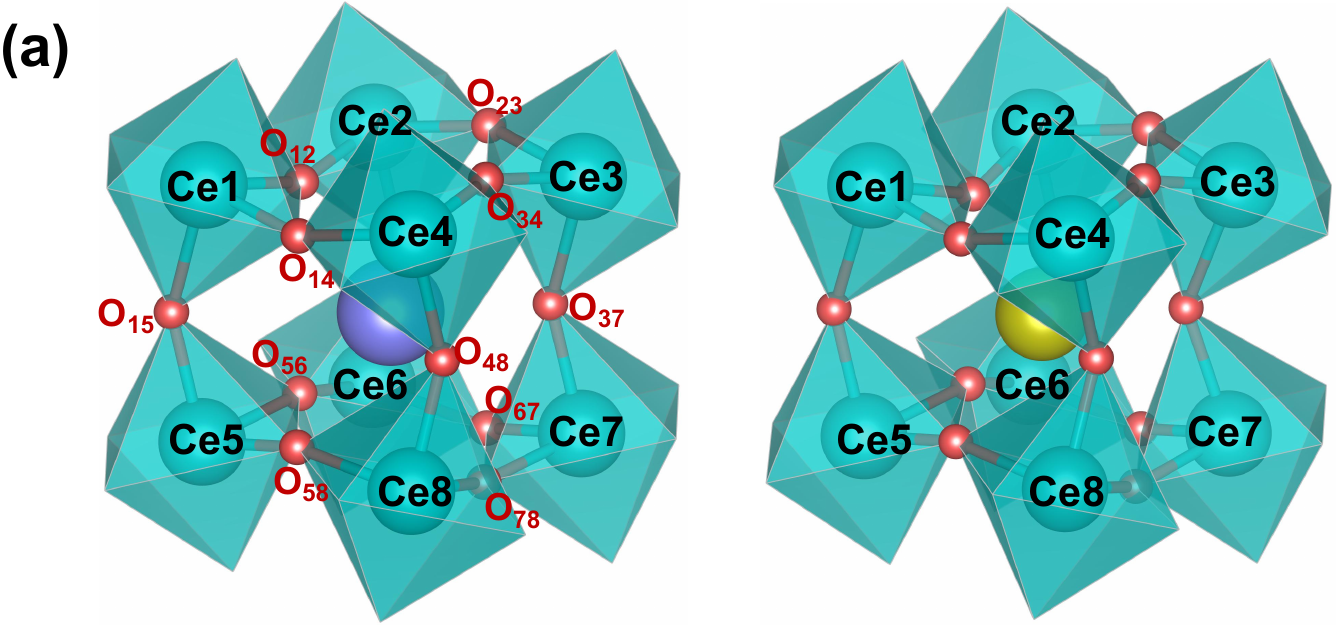}\\[0.02cm]
\includegraphics[width=0.45\textwidth]{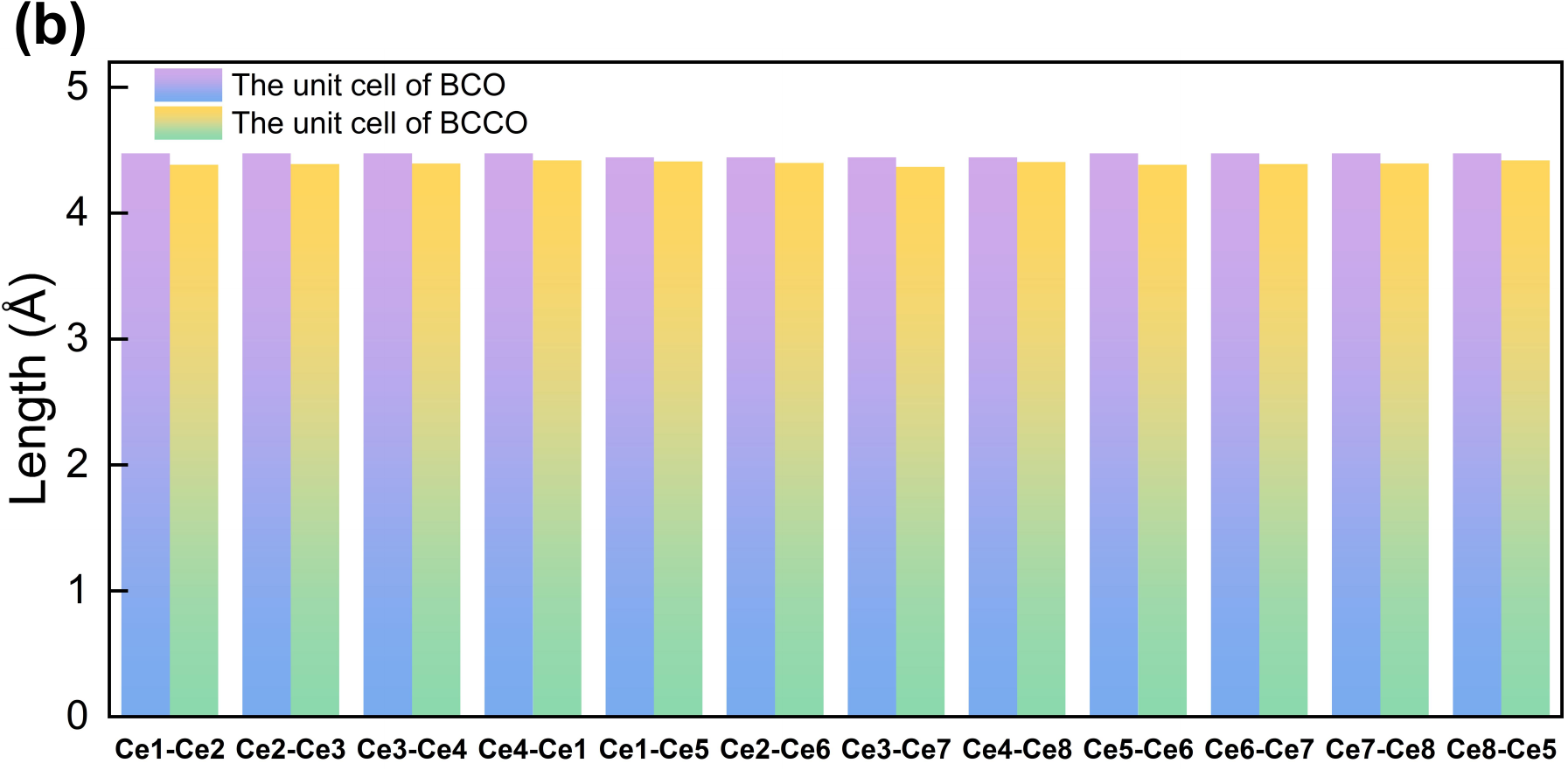}\\[0.3cm]
\includegraphics[width=0.45\textwidth]{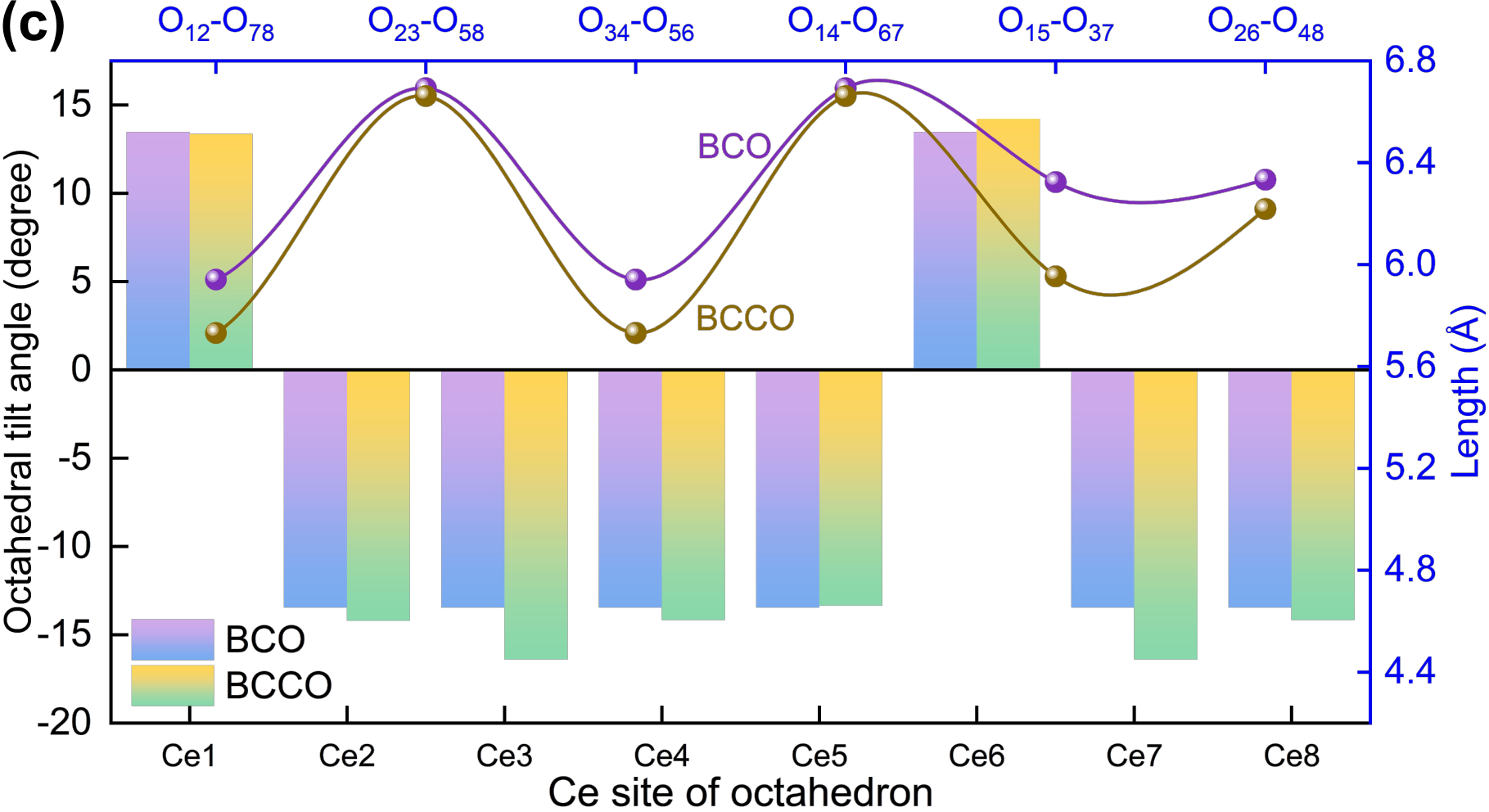}
\caption{(a) The left panel shows the unit cell containing a Ba atom in the BCO supercell, and the right panel shows the unit cell containing a Ca atom in the BCCO supercell. The corner Ce atoms and their octahedra are labeled as Ce1-Ce8. The subscripts of the oxygen atoms located on the cell edges correspond to the two Ce atoms'label they are bonded to, such as O$_{12}$-O$_{78}$ shown in the figure. (b) Measurements of the cell edge lengths for the Ba-containing and Ca-containing unit cells corresponding to panel (a). (c) Measurements of the octahedral tilt angles and the distances between the farthest diagonal O atoms corresponding to panel (a). The octahedral rotation around the z direction is defined as positive when directed into the unit cell and negative when directed out of it. }
\label{Fig2}
\end{figure}

\begin{figure}[htbp]
\centering
\includegraphics[scale=0.53]{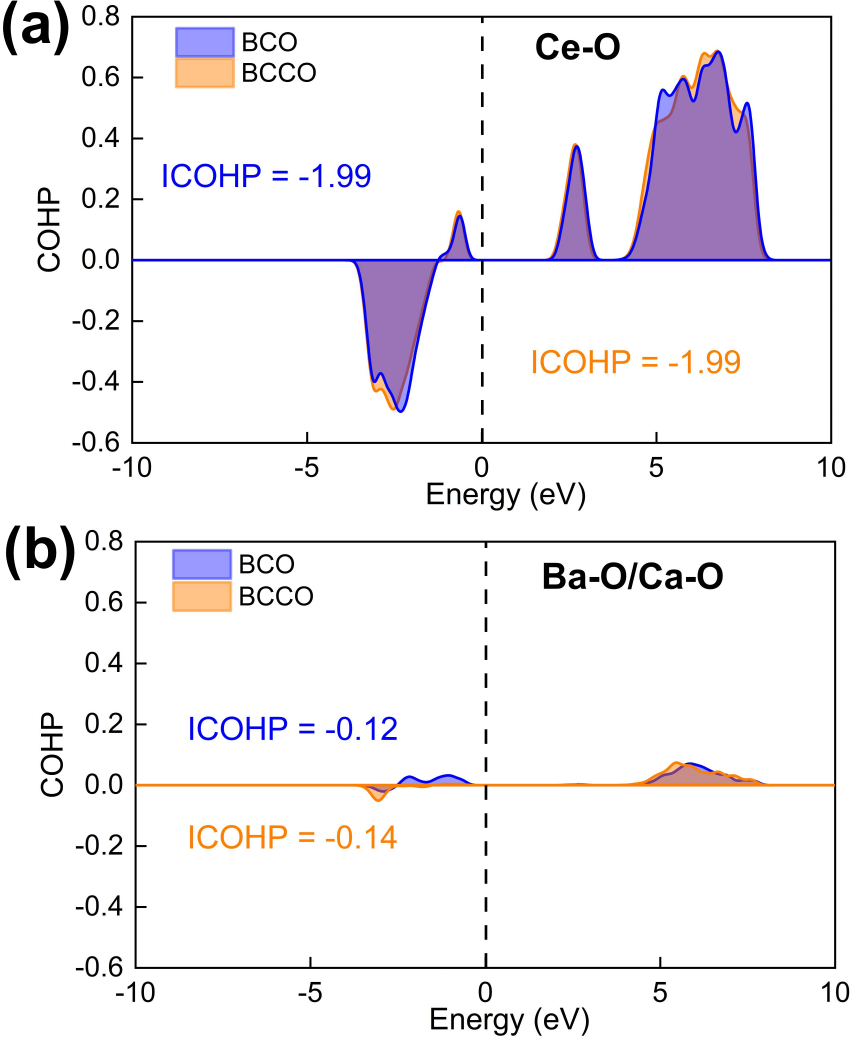}
\caption{(a)  The COHP and ICOHP of Ce-O bonds in BCO and BCCO. (b) The COHP and ICOHP of Ba-O bonds in BCO and Ca-O bonds in BCCO. The plotted values are obtained by averaging over all Ce-O and Ba-O bonds within one Ba-containing unit cell in the BCO supercell, and over all Ce-O and Ca-O bonds within one Ca-containing unit cell in the BCCO supercell.}
\label{Fig3}
\end{figure}

We employed the orthorhombic phase of BCO primitive cell, which is stable at room temperature\cite{jacobson1972structures}. The structure belongs to the Pnma space group, and according to the Materials Project database, its energy above hull is 0 eV/atom. The optimal lattice constants of primitive cell are a=6.35~\AA, b=6.33~\AA, c=8.81~\AA, which are in good agreement with experimental values (a=6.25~\AA, b=6.23~\AA, c=8.79~\AA)\cite{knight1994structural}. The slight overestimation of the lattice constants is consistent with the general behavior of PBE exchange-correlation functional calculations. After correcting for systematic deviations, the residual deviation is within 1.1\%\cite{lejaeghere2014error}, confirming the reliability of the computational parameters. A $\sqrt{2} \times \sqrt{2} \times 1$ supercell was constructed to model the bulk structure. The optimized lattice parameter of the pristine BCO supercell is a=8.95~\AA, b=8.95~\AA, c=8.89~\AA , which decreases to a=8.88~\AA, b=8.90~\AA, c=8.89~\AA\ after substituting one Ba atom with Ca. The reduction in lattice volume is mainly attributed to the lattice contraction effect. As shown in Fig.~\ref{Fig2}(a), the eight Ce atoms located at the corners of the primitive cell are labeled Ce1-Ce8. We measured the distances between adjacent Ce ions to evaluate the cell edge lengths. After substituting Ba with Ca, all edge lengths decrease (Fig.~\ref{Fig2}(b)), indicating a contraction of the Ce framework within the cell. Meanwhile, the distances between the most distant diagonal oxygen pairs are also reduced, as shown in Fig.~\ref{Fig2}(c), reflecting a decrease in the free volume of the unit cell. This observed lattice contraction originates from the smaller ionic radius of Ca$^{2+}$ ($R_{\mathrm{Ca}^{2+}}=1.34$ \AA) compared with Ba$^{2+}$ ($R_{\mathrm{Ba}^{2+}}=1.61$ \AA), which reduces the Goldschmidt tolerance factor of the system (from 0.94 to 0.85 upon Ca substitution). As a result, the CeO$_6$ octahedra undergo stronger tilting to fill the free space at the A site, alleviating the mismatch between the A-O and B-O bond. And the COHP analysis in the Fig.~\ref{Fig3}(a) indicates that the Ce-O bond length is essentially unchanged upon Ca substitution. Therefore, the enhanced CeO$_6$ octahedral tilting reduces the distance between neighboring Ce atoms forming the lattice framework by decreasing the Ce-O-Ce bond angle further away from 180$^\circ$.

Figure~\ref{Fig3} presents the Crystal Orbital Hamilton Population (COHP)of the BCO and BCCO systems obtained using the LOBSTER code\cite{maintz2016lobster}. In the BCO structure, the Ce-O bonds exhibit distinct antibonding and bonding peaks above and below the Fermi level, and the integrated COHP (ICOHP) value is -1.99, indicating a typical covalent bonding nature of Ce-O interactions\cite{liu2020microscopic}. In contrast, the COHP of the Ba-O bonds shows no clear separation between bonding and antibonding regions and is dominated by antibonding states. Its peak value is much smaller than that of the covalent-bond-dominated semiconductor CaAs (2.5)\cite{nelson2020lobster}, suggesting weak and energetically unfavorable orbital overlap. Therefore, the Ba-O bond is primarily ionic in nature. Upon Ca doping, as shown in Fig.~\ref{Fig3}(a), the COHP and ICOHP of the Ce-O bonds in BCCO exhibit no significant change, indicating that the covalent strength of the Ce-O bonds remains nearly unchanged. Meanwhile, the Ca-O bonds show a weak bonding state below the Fermi level, and the ICOHP value (0.14) is slightly larger than that of the Ba-O bonds (0.12), suggesting a slightly enhanced covalent contribution. Nevertheless, the small ICOHP value indicates that the orbital overlap remains weak, and the Ca-O bonds are still predominantly ionic in nature. This can also be seen from the projected density of states (PDOS) in Fig.~\ref{FigA1}, where the Ba-O/Ca-O bonds exhibit much weaker orbital overlap at deeper energy levels compared to the Ce-O bonds. Therefore, in the following analysis, the Bader charge method\cite{henkelman2006fast}, which partitions atoms based on the charge density, is employed to examine the strength of Ba-O/Ca-O bonds. And the higher electronegativity of Ca relative to Ba reduces the degree of charge transfer to oxygen, resulting in weaker A-O ionic bonding. Figures~\ref{Fig1}(c) and 1(d) show the band structures and DOS of BCO and BCCO, respectively. The undoped BCO exhibits a band gap of 2.8 eV, confirming its insulating nature, which is consistent with previous theoretical reports\cite{he2021new}. After Ca doping, no significant changes are observed in either the band structure or the DOS especially near the Fermi level, indicating that the electronic structure remains nearly unaffected. Therefore, the influence of lattice distortion on proton conduction can be investigated independently. Moreover, the insulating nature of BCO and BCCO is beneficial for maintaining a high open-circuit voltage\cite{zhou2016strongly}, ensuring efficient SOFC performance.

\subsection*{B. Oxygen Vacany Formation Energy of Bulk and Surface}
\begin{figure*}[htbp]
\centering
\includegraphics[scale=0.5]{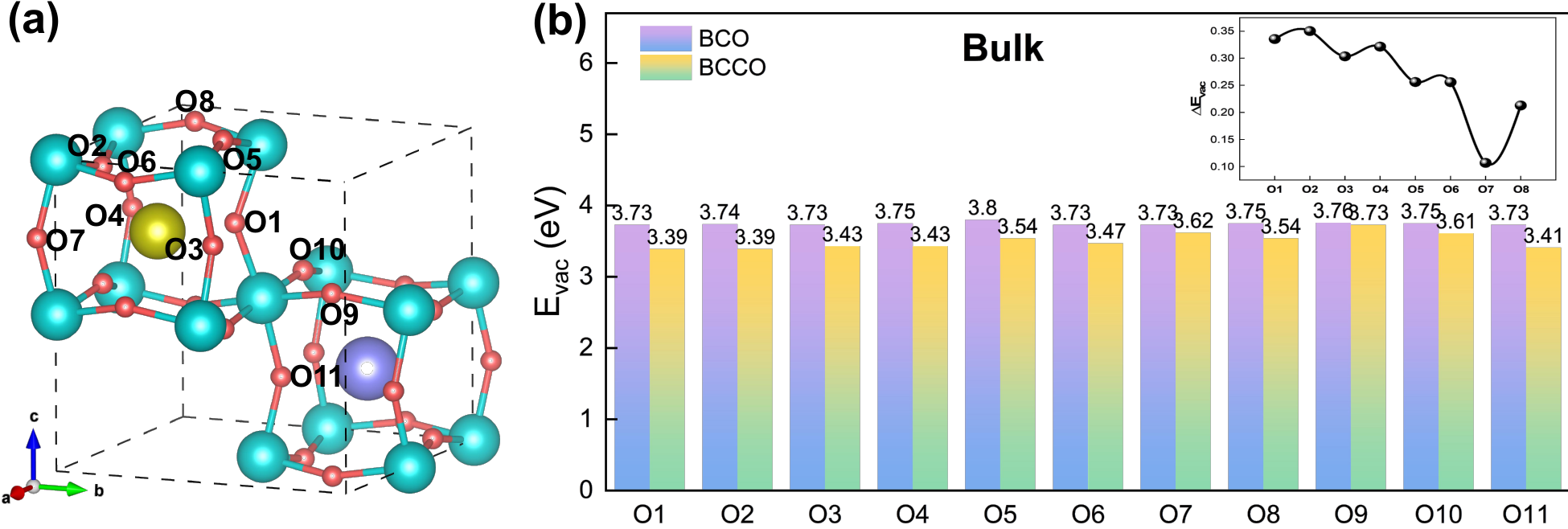}
\caption{(a)  Schematic illustration of the 11 inequivalent oxygen sites (O1-O11) considered for calculating the oxygen vacancy formation energies in the bulk BCCO structure. (b) Oxygen vacancy formation energies of O1-O11 in BCO and BCCO. The inset shows the difference in formation energy ($\delta E_{vac}$) at the nearest-neighbor oxygen sites (O1-O8) between the two systems.}
\label{Fig4}
\end{figure*}

To explain the experimentally observed enhancement of proton conductivity in Ca-doped BaCeO$_3$-based materials\cite{dudek2016some,luo2023chemical,kothandan2024effect}, we calculated the oxygen vacancy formation energies ($E_{vac}$) of the bulk and surface structures of BCO and BCCO, which serve as important indicators of proton concentration\cite{liu2020microscopic,loken2016alkali,kruth2003water}. The formula is\cite{he2022surface,zhang2018insight}:
\begin{equation}
E_{vac} = E(V_O^{\bullet\bullet})-E_{perfect}+\frac{1}{2}E_{O_2}
\label{Equ1}
\end{equation}
Here, $E(V_O^{\bullet\bullet})$ and $E_{perfect}$ denote the total energies of the supercells with and without an oxygen vacancy, respectively. $E_{O_2}$ represents the energy of an oxygen molecule, obtained from a density functional theory calculation of a single O$_2$ molecule placed in a 15~\AA\ cubic box.

We first examine the formation of oxygen vacancies in the bulk structure. We calculated the oxygen vacancy formation energies $E_{vac}$ for the inequivalent oxygen sites O1-O8 located within the unit cell containing Ca, as well as for the sites O9-O11 in the neighboring unit cell where the dopant has no direct influence [Fig.~\ref{Fig4}(a)], with the labels O1-O11 assigned in order of increasing distance from the Ca atom. The former are collectively referred to as the nearest-neighbor oxygen sites of Ca, and the latter as the next-nearest-neighbor sites. The results are shown in Fig.\ref{Fig4}(b). For the BCO system, according to the space group and bonding environment analysis, O2, O5, O6, O8, O9, and O10, which are located along the ab-plane of the octahedron, correspond to the equivalent 8d O, while O1, O3, O4, O7, and O11, positioned along the c-axis, correspond to the equivalent 4c O. Therefore, the $E_{vac}$ of these equivalent oxygen sites are nearly identical. Upon Ca doping, however, the structural symmetry is broken, and these oxygen atoms are no longer equivalent, leading to slight differences in their $E_{vac}$ values. Nevertheless, all oxygen sites exhibit a decrease in $E_{vac}$ after Ca substitution. For the nearest-neighbor oxygen sites (O1-O8), the decrease in $E_{vac}$ becomes more pronounced as the distance from Ca decreases (see the inset of Fig.~\ref{Fig4}(b)). These results indicate that A-site Ca substitution in the BaCeO$_3$ bulk favors the formation of oxygen vacancies and thus facilitates proton incorporation. It should be noted that oxygen vacancies in proton-conducting electrolyte materials are generated through the charge-compensation mechanism associated with acceptor doping. Therefore, we constructed a supercell of the original 40-atom structure and introduced two Y dopants at the B site, which ensures charge neutrality upon the formation of one oxygen vacancy. The calculated oxygen-vacancy formation energy decreases by 0.31 eV after Ca doping, in agreement with the result shown in Fig.~\ref{Fig4}(b). Therefore, Eq.~\ref{Equ1} can be used to assess the effect of A-site Ca doping on the thermodynamic tendency for oxygen-vacancy formation.

\begin{figure}[htbp]
\centering
\includegraphics[scale=0.43]{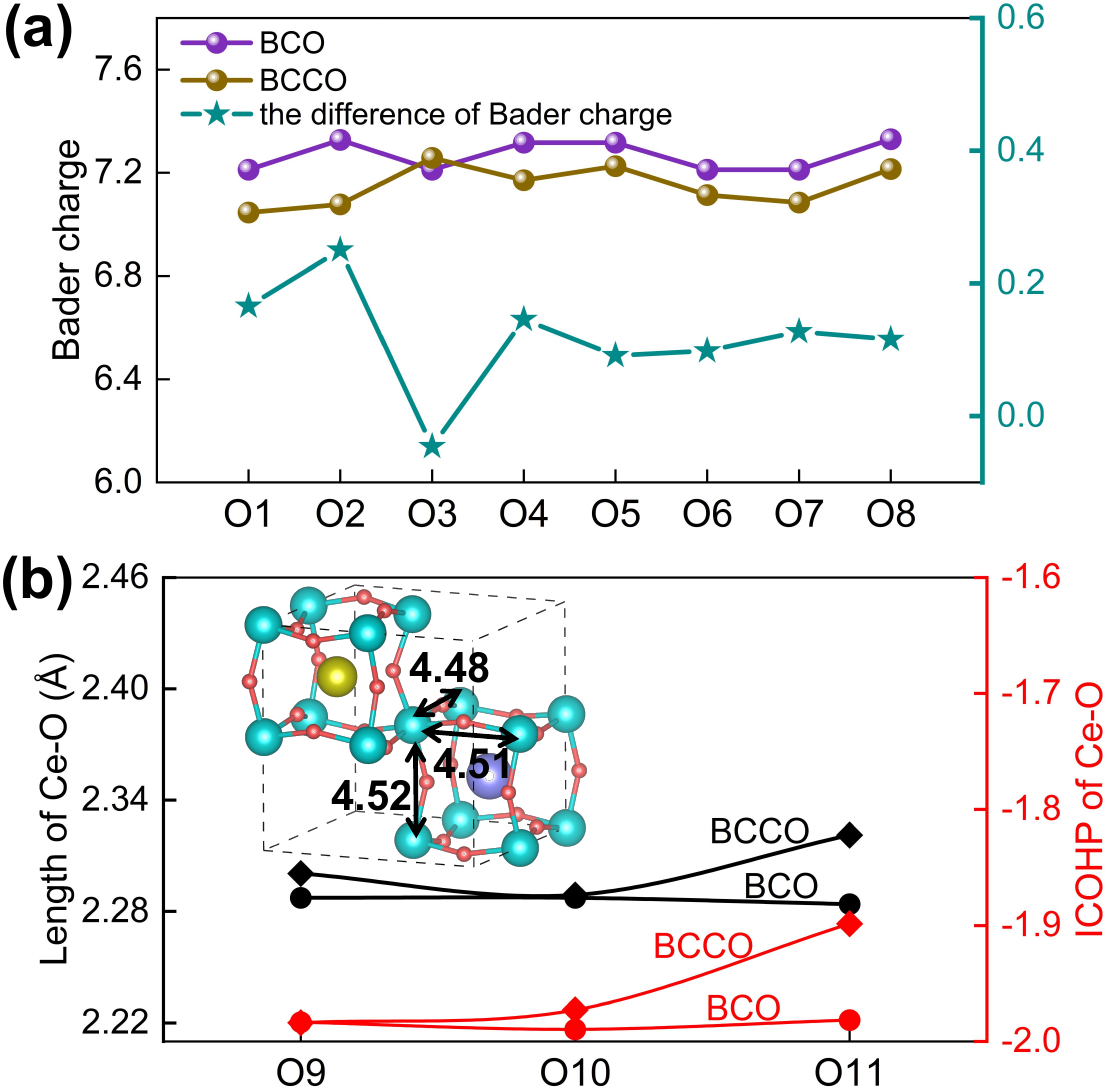}
\caption{(a)  Bader charges of the oxygen sites O1-O8 in BCO and BCCO, together with their differences between the two systems. (b) Averaged Ce-O bond lengths and ICOHP for O9-O11 (each oxygen bonded to two Ce atoms). The inset shows the schematic illustration of the unit-cell edge containing O9-O11.}
\label{Fig5}
\end{figure}

To elucidate why the doping of the isovalent ion Ca leads to a decrease in the oxygen vacancy formation energy, we calculated the Bader charges of the O1-O8 sites, as shown in Fig.~\ref{Fig5}(a). The oxygen atoms surrounding Ca exhibit lower Bader charges than those near Ba (except for O3), indicating  weaker Ca-O ionic bonding. Meanwhile, the Ce-O covalent bond strength remains nearly unchanged(Fig.~\ref{Fig3}(a)). Consequently, oxygen vacancies are more easily formed. Moreover, as shown in Fig.~\ref{Fig5}(a), the variation in the charge difference between the oxygen atoms surrounding Ca and those around Ba follows the same trend as the difference in oxygen vacancy formation energies illustrated in the inset of Fig.~\ref{Fig4}(b)-the closer the oxygen site is to the A-site ion, the stronger the effect. This further confirms that the change in A-O ionic bond strength is responsible for the variation in vacancy formation energies at the nearest-neighbor oxygen sites. The inconsistency observed for O3 may arise from the zero-flux surface partitioning method used in the Bader charge analysis, which tends to overestimate the charge of O3 near Ca\cite{henkelman2006fast}.

As discussed in Section A, the incorporation of Ca leads to a local lattice contraction in the substituted unit cell, which in turn indirectly affects the oxygen vacancy formation energies at the next-nearest oxygen sites (O9-O11). As shown in Fig.~\ref{Fig5}(b), upon introducing one Ca atom into the BCO supercell, the edge lengths of the Ce framework in the neighboring Ba-centered cell (measured along three directions) increase from the original 4.48~\AA\ to 4.48~\AA, 4.51~\AA, and 4.52~\AA, respectively. Consequently, the average Ce-O bond lengths associated with O9, O10, and O11 become longer, accompanied by a reduction in the corresponding ICOHP magnitudes, indicating weakened Ce-O bonding interactions. The reduced bond thereby decreases the $E_{vac}$. These results suggest that A-site Ca doping indirectly promotes the formation of oxygen vacancies in the neighboring expanded cells through the induced lattice contraction effect.

\begin{figure}[htbp]
\centering
\includegraphics[scale=0.36]{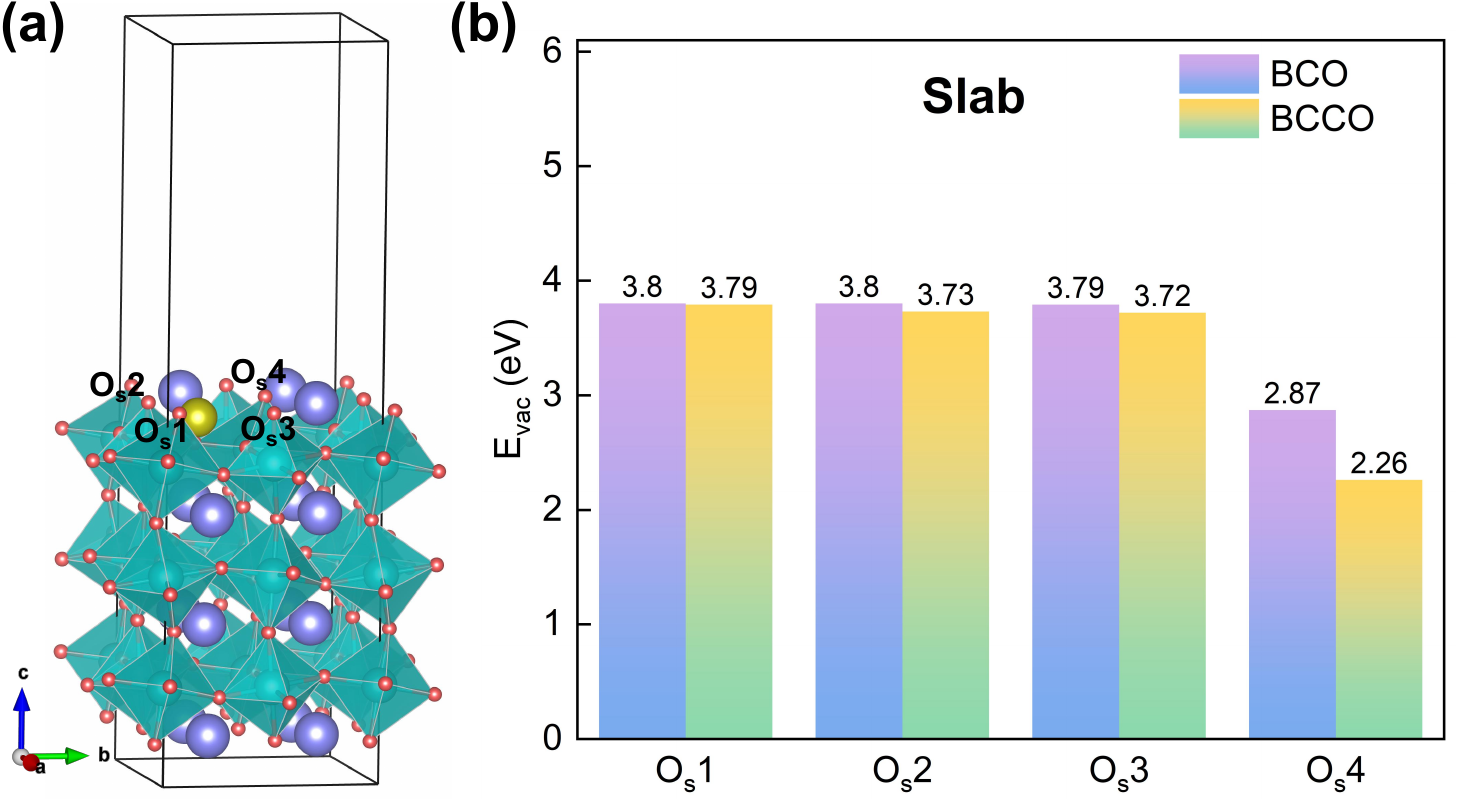}
\caption{(a) Illustration of the oxygen sites O$_s$1-O$_s$4 in the BCCO slab used for calculating the oxygen vacancy formation energies. (b) $E_{vac}$ of O$_s$1-O$_s$4 in the BCO and BCCO slab systems.}
\label{Fig6}
\end{figure}

Because the Ca-O bond is weaker than the Ba-O bond, Ca tends to segregate from the bulk to the surface, where the coordination number is reduced, in order to maximize the number of stronger Ba-O bonds\cite{tauer2013computational} and thereby lower the total energy of the system. As shown in Fig.~\ref{FigA2}, the total energies of the slab with Ca located in the bulk and on the surface are -480.37 eV and -480.83 eV, respectively. Therefore, in the following, we analyze the effect of Ca segregation to the surface on the oxygen vacancy formation. As shown in Fig.~\ref{Fig6}, we calculated the oxygen vacancy formation energies for the four oxygen sites (O$_s$1-O$_s$4) surrounding Ca. It can be seen that, after Ca substitutes for Ba at the surface, all oxygen sites except O$_s$4 exhibit a slightly lower $E_{vac}$ compared with the bulk BCCO case, and the absolute values of $E_{vac}$ remain higher than those in the bulk shown in Fig.~\ref{Fig4}(b). This indicates that the promotion effect of Ca on oxygen vacancy formation is not significant at the surface, which helps suppress the accumulation of positively charged oxygen vacancies at interfaces and thereby mitigates grain-boundary resistance\cite{tauer2013computational}, favoring proton conduction.

Compared with the bulk, this weaker promotion of oxygen vacancy formation mainly originates from structural reconstruction, which arises from lattice distortions caused by the reduced atomic coordination at the surface as the system relaxes to restore local energetic stability\cite{hutner2024stoichiometric}. As shown in Fig.~\ref{FigA3}, the (001) surface of the slab model exhibits a stronger octahedral tilting than that of the bulk, which shortens the Ca-O distances (Ca-O$_s$1 and Ca-O$_s$2) from 2.49 and 2.67~\AA, to 2.27 and 2.33~\AA, respectively. Consequently, the covalent character of the Ca-O bonds is enhanced, with the ICOHP values of Ca-O$_s$1 and Ca-O$_s$2 being -0.32 and -0.29, larger in magnitude than the bulk value of -0.14 (Fig.~\ref{Fig3}(b)), indicating stronger bonding. The Ce-O bonds at the surface are likewise shortened: the Ce-O$_s$1, Ce-O$_s$2, Ce-O$_s$3, and Ce-O$_s$4 bond lengths decrease from 2.27, 2.27, 2.28, and 2.32~\AA\ to 2.21, 2.21, 2.19, and 2.19~\AA, respectively. As a result, their ICOHP values (-2.36, -2.35, -2.42, and -2.40) are all larger in magnitude than the bulk value of -1.99 (Fig.~\ref{Fig3}(a)), confirming enhanced Ce-O bonding. These strengthened Ca-O and Ce-O bonds make the oxygen vacancy formation energy of BCCO at the surface higher than that in the bulk, and no obvious reduction is observed compared with BCO. However, the $E_{vac}$ at the O$_s$4 site shows an opposite trend. This is mainly because the structural reconstruction induced by Ca at the surface elongates the Ca-O$_s$4 distance, thereby weakening the electrostatic interaction between Ca and O. As a result, the ionic bond strength is reduced, and this reduction effect exceeds the enhancement of the Ce-O$_s$4 bond strength, leading to a lower oxygen vacancy formation energy at O$_s$4. Nevertheless, the relatively high vacancy formation energies at most sites help to avoid the accumulation of the grain-boundary core charge.

\subsection*{C.Hydration Reaction}
\begin{figure}[htbp]
\raggedright
\includegraphics[scale=0.37]{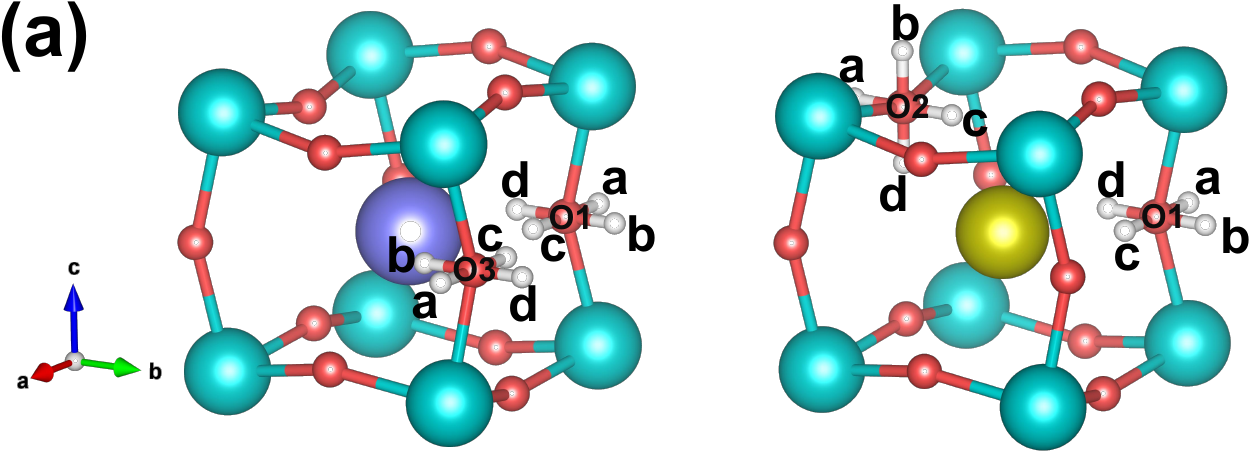}\\
\includegraphics[scale=0.5]{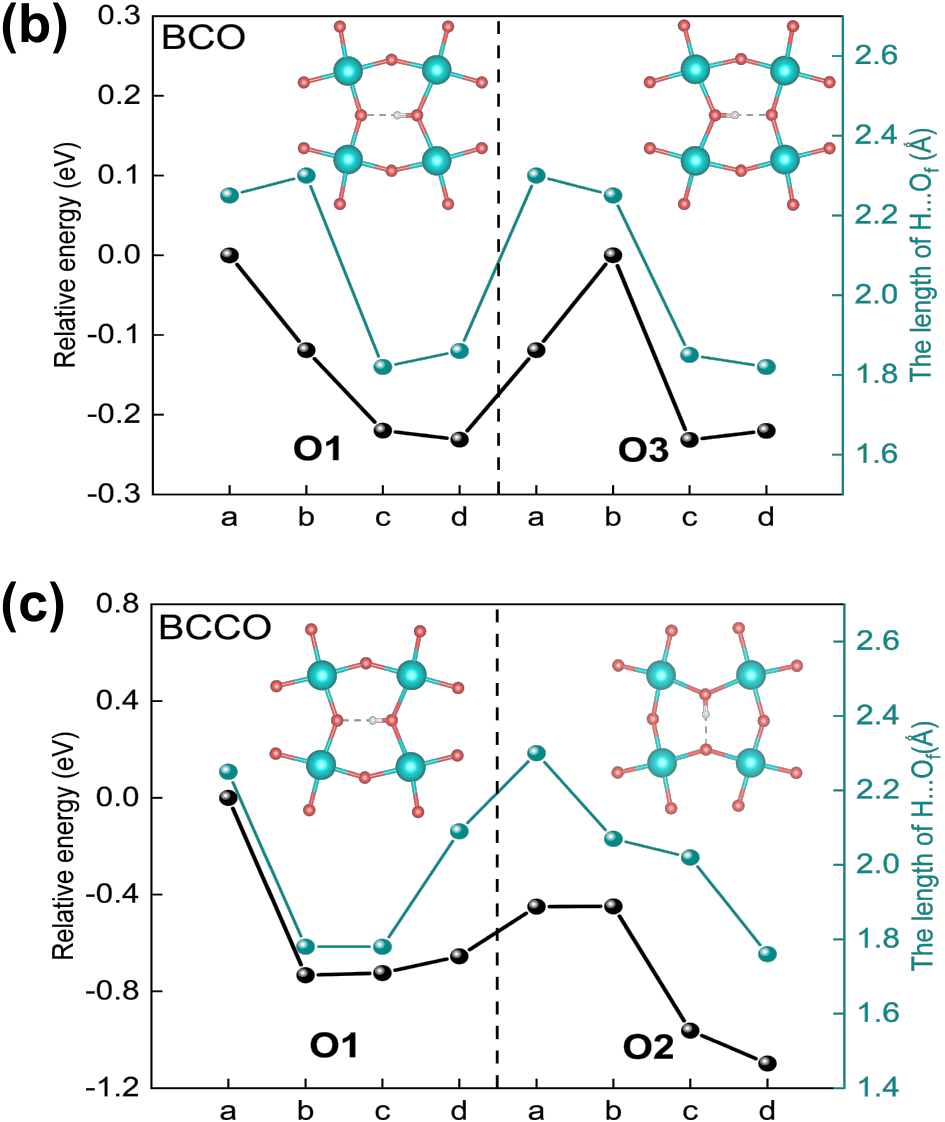}
\caption{(a) Illustrations of different proton orientations at the selected O1 and O3 sites in the BCO system (left), and at the selected O1 and O2 sites in the BCCO system (right). (b) Relative energies and corresponding H...O$_f$ hydrogen bond lengths for different proton configurations in BCO, corresponding to the left panel of (a). (c) Relative energies and H...O$_f$ lengths for different proton configurations in BCCO, corresponding to the right panel of (a). The insets depict the lowest-energy proton orientations at the respective oxygen sites. The relative energies are referenced to the proton configuration with the highest energy.}
\label{Fig7}
\end{figure}

The generated oxygen vacancies can promote the dissociation of water, enabling proton incorporation\cite{domingo2019water}. Therefore, in this part, we calculate the hydration energy ($\Delta E_{hyd}$) to evaluate how Ca doping affects the proton uptake ability of BaCeO$_3$-based materials\cite{loken2016alkali}. According to the hydration reaction: $H_2O + V_O^{\bullet\bullet} + O_O^{\times} \rightarrow 2OH_O^{\bullet}$, the hydration energy $\Delta E_{hyd}$ can be expressed as \cite{he2022surface,dawson2015first}:
\begin{multline}
\Delta E_{hyd} = 2E_{tot}(hydrated)-E_{tot}(V_O^{\bullet\bullet}(1))-E_{tot}(V_O^{\bullet\bullet}(2))\\-2E_{tot}(H_2O)
\label{Equ2}
\end{multline}
Here, $E_{tot}(hydrated)$ is the total energy of the hydrated system, while $E_{tot}(V_O^{\bullet\bullet}(1))$ and $E_{tot}(V_O^{\bullet\bullet}(2))$ represent the total energies of systems containing one oxygen vacancy each. Two different oxygen-vacancy configurations are considered because the hydration reaction involves two oxygen sites that bind protons. $E_{tot}(H_2O)$ denotes the total energy of a single water molecule, which is obtained by optimizing the structure of an isolated H$_2$O molecule placed in a 15 \AA\ cubic box.

Before calculating the hydration energy, it is necessary to investigate the energies of different proton configurations in the BCO and BCCO. To identify the most probable proton sites, we selected two oxygen sites with the lowest oxygen vacancy formation energies (Fig.~\ref{Fig4})-- O1 and O3 in BCO, O1 and O2 in BCCO. For each oxygen site, four possible proton orientations (a, b, c, d) were considered, as illustrated in Fig.~\ref{Fig7}(a). The consistent trend between the energies of different proton configurations and the length of the formed hydrogen bonds indicates that the low energies of the configurations (O1d and O3c in BCO, O1b and O2d in BCCO) originate from the formation of strong attractive hydrogen bonds between the proton and its neighboring oxygen atoms. The reason why the protons in these configurations form the shortest and strongest hydrogen bonds is that the corresponding oxygen sites are bent inward toward the proton orientation, thereby shortening the proton-oxygen distance (as illustrated in the insets of Figs.~\ref{Fig7}(b) and \ref{Fig7}(c)). These lowest-energy proton orientations were selected for the calculation of $\Delta E_{hyd}$.

Using Eq.~\ref{Equ2}, the calculated hydration energies are -2.41 eV for BCO and -3.03 eV for BCCO, indicating that Ca doping also promotes the hydration reaction. We further calculated the hydration energies in the oxygen-vacancy environment generated by acceptor Y doping. The hydration energies before and after Ca doping are -2.83 eV and -3.53 eV, respectively, corresponding to a reduction of 0.70~eV, which is consistent with the above results. This may be related to the enhanced basicity of the system after Ca doping\cite{dawson2015first,kreuer1999aspects,kreuer2001proton}. As shown in Fig.~\ref{FigA4}, the charge density around Ca is significantly higher than that around Ba at the same reference charge density level. The calculated Bader charge of Ca is 8.55, larger than that of Ba 8.33 (10 valence electrons in the pseudopotential), which is associated with the weakened Ca-O ionic bond and reduced charge transfer. Consequently, after Ca doping. the A-site cations become more electron rich (more basic), thereby enhancing the ability of the material to absorb protons and promoting the hydration reaction. It is worth noting that Ce exhibits a certain degree of multivalency and thus has the potential for spontaneous hydrogen incorporation\cite{zhou2016strongly,islam2020computational,linghu2025multivalent}. We therefore calculated the hydrogen insertion energy in stoichiometric BaCeO$_3$ and obtained a value of 0.21 eV. The positive value indicates that BaCeO$_3$, as a conventional proton-conducting electrolyte, primarily incorporates protons via the oxygen-vacancy-mediated hydration mechanism described by Eq.~\ref{Equ2}.

\subsection*{D. Proton Diffusion in BCO and BCCO}
\begin{figure}[htbp]
\raggedright
\includegraphics[width=0.4\textwidth]{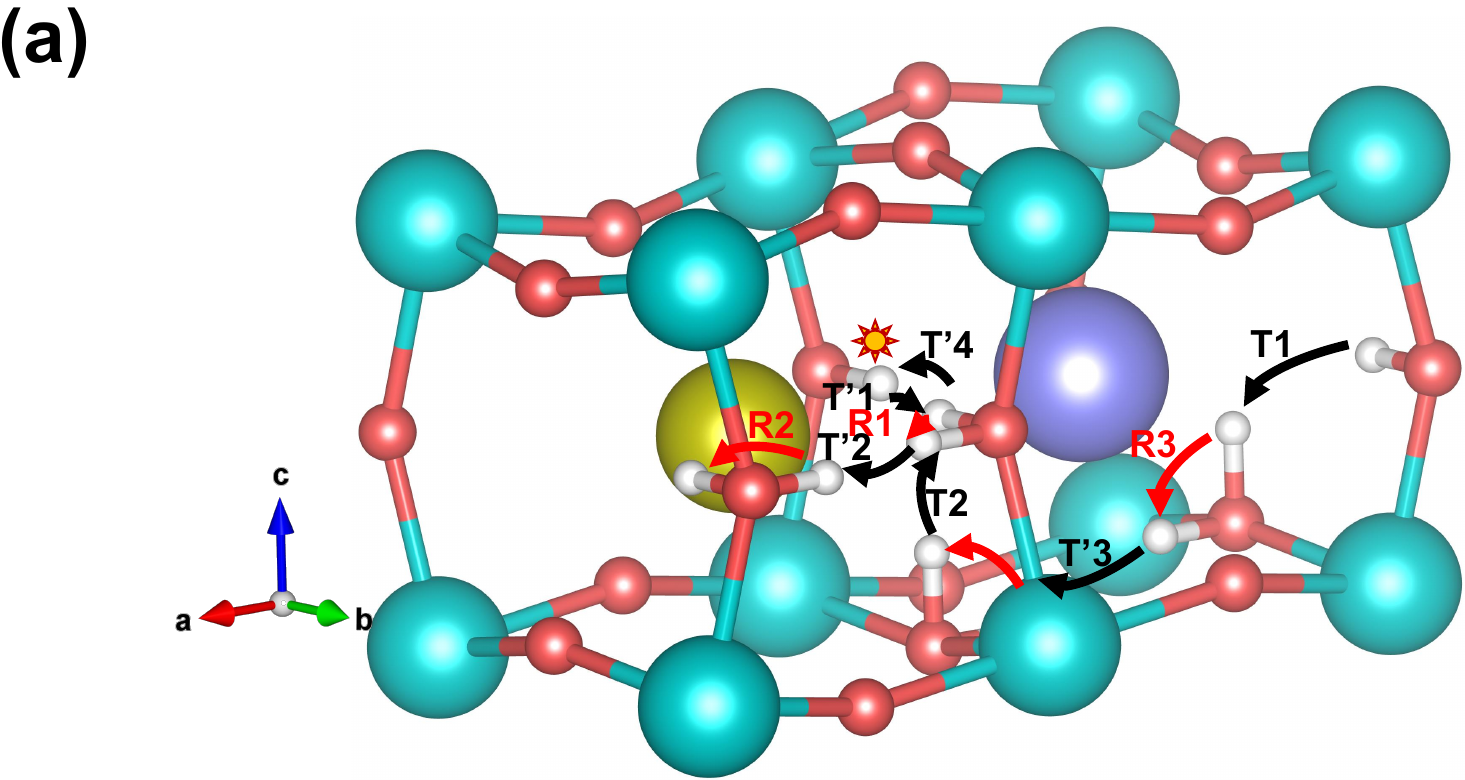}\\[0.05cm]
\includegraphics[width=0.5\textwidth]{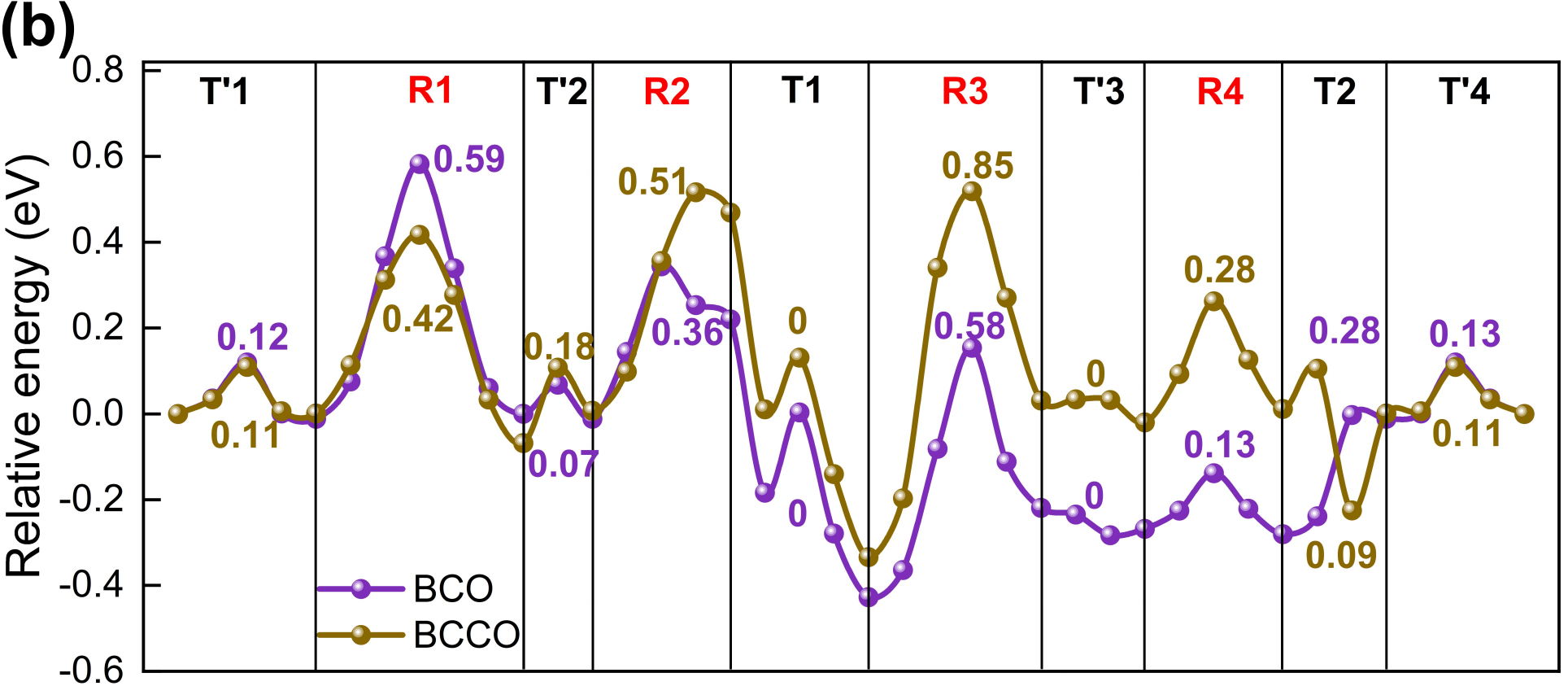}
\caption{(a) A periodic minimum-energy migration pathway in the BCCO system. The sun symbols mark the starting and ending points of the path. Proton transfer (T and T$'$) are indicated by black arrows, and rotation (R) by red arrows. (b) Energy barriers for each migration step along the pathway in (a) for the BCO and BCCO supercells. The local minima correspond to the stable proton sites along the complete diffusion pathway shown in panel (a).}
\label{Fig9}
\end{figure}

In addition to proton concentration, proton diffusion is also a key factor determining proton conductivity\cite{liu2020microscopic,PhysRevB.82.014103}. The proton diffusion follows the Grotthuss mechanism\cite{agmon1995grotthuss,kreuer2000complexity}, which involves two elementary steps: the rotation of the proton around a single oxygen site and the transfer between different oxygen sites. The latter can be further classified into intraoctahedral transfer between neighboring oxygens within the same octahedron and interoctahedral transfer between oxygens belonging to adjacent octahedron \cite{merinov2009proton}. To evaluate the effect of Ca doping on the proton diffusion capability of the BCO system, we calculated a periodic minimum-energy migration pathway: $T'1 \rightarrow R1 \rightarrow T'2 \rightarrow R2 \rightarrow T1 \rightarrow R3 \rightarrow T'3 \rightarrow R4 \rightarrow T2 \rightarrow T'4$, which consists of ten alternating processes of rotation (R), intraoctahedral transfer (T), and interoctahedral transfer (T$'$). After each transfer or rotation event, the proton moves to the adjacent stable oxygen site indicated by the arrows in Fig.~\ref{Fig9}(a). Compared with the BCO system, most rotation barriers in the BCCO system are significantly increased, whereas the barriers for two types of transfer processes exhibit slight decreases (Fig.~\ref{Fig9}(b)). For both BCO and BCCO, most rotation processes have much higher energy barriers than the transfer processes. For example, the maximum rotational barrier (R1) in BCO is 0.59 eV, whereas the maximum transfer barrier (T2) is only 0.28 eV. Since the proton migration rate is exponentially negatively correlated with the migration barrier\cite{liu2020microscopic,PhysRevB.82.014103,lin2024comparative}, the 0.31 eV difference between the two processes is sufficient to cause more than an order of magnitude difference in migration rates at intermediate operating temperatures \cite{ma2026hydrogen,PhysRevB.87.104303}. Therefore, the rotation process serves as the rate-limiting step, which the proton must undergo for long-range diffusion. And Ca doping increases this rate-limiting barrier, thus hindering long-range proton diffusion. This further indicates that the enhanced proton conductivity in the Ca-doped BaCeO$_3$-based system mainly originates from the increased proton concentration caused by the easier formation of oxygen vacancies and hydration reactions.

\begin{figure}[htbp]
\centering
\includegraphics[scale=0.45]{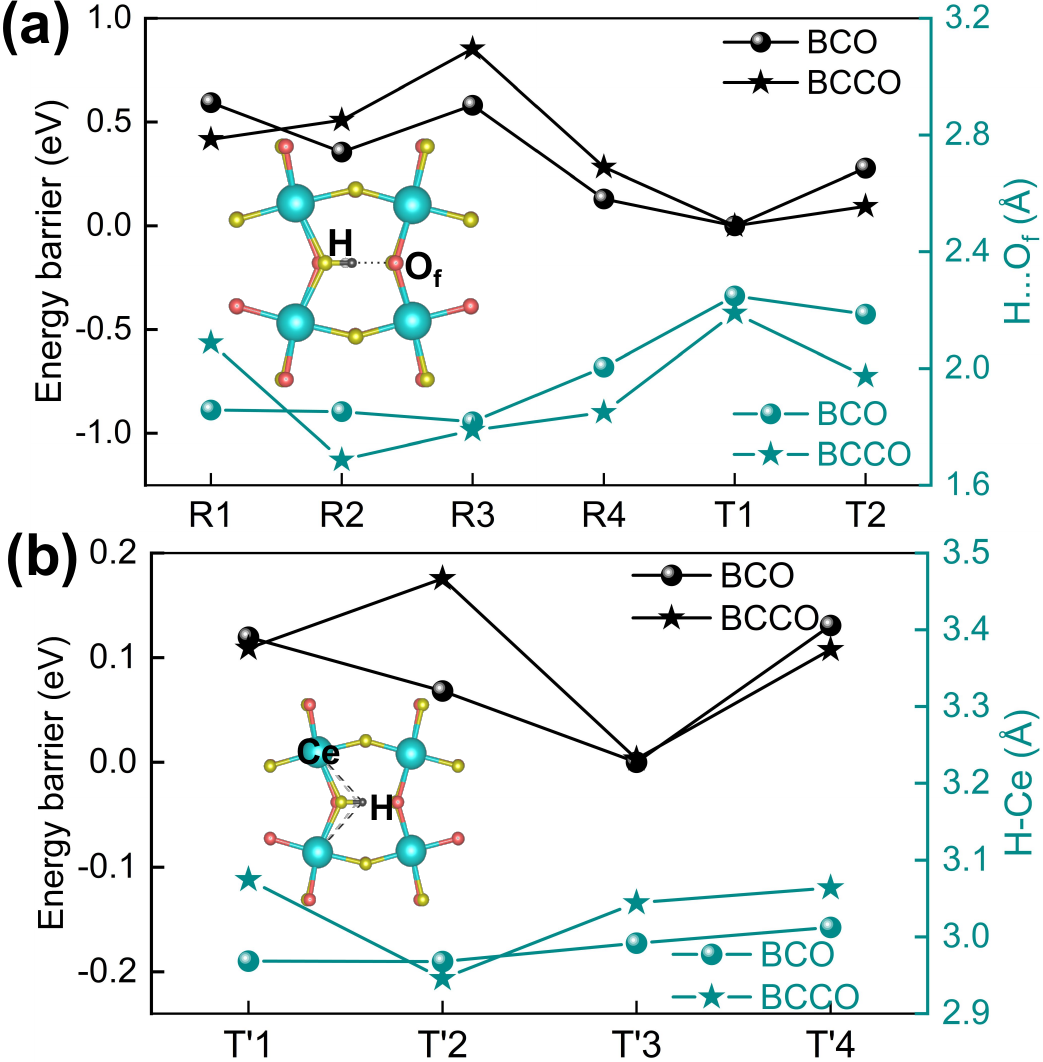}
\caption{(a)Comparison of the energy barriers for all rotation (R) and intraoctahedral transfer (T) processes along the migration path shown in Fig.~\ref{Fig9}(a), together with the corresponding initial H...O$_f$ hydrogen bond lengths in BCO and BCCO. The inset illustrates the lattice contraction effect that shortens the hydrogen bond, which the yellow and black spheres represent the oxygen and hydrogen atoms after lattice contraction, respectively. (b) Comparison of the energy barriers for all interoctahedral transfer (T$'$) processes and the initial H-Ce distances in BCO and BCCO. The inset illustrates the lattice contraction effect that elongates the H-Ce distance.}
\label{Fig10}
\end{figure}

According to the proton-lattice coupling mechanism\cite{du2020cooperative,samgin2000lattice,kreuer1996proton,jing2020role}, both the transfer and rotation processes are strongly influenced by the structural descriptors, including the initial hydrogen-bond length H...O$_f$ and the distances between the proton and neighboring A-site or B-site cations\cite{munch1997quantum,MUNCH2000183,D2TA08664F,doi:10.1021/acs.chemmater.1c02432,D4EE01219D,merinov2009proton,islam2022first,chung2026flexibility}. Therefore, we analyze the bond length variations before and after Ca doping to elucidate how the resulting lattice contraction affects the proton migration barrier. As shown in Fig.~\ref{Fig10}(a), the consistent trend between the rotation and intraoctahedral transfer barriers and the initial H...O$_f$ hydrogen bond length in both BCO and BCCO indicates that, due to the enhanced octahedral tilting induced by Ca doping, the hydrogen bond between the proton and the acceptor oxygen becomes shorter(the insert of Fig.~\ref{Fig10}(a)). This shortening makes it more difficult for the lattice bending to break the hydrogen bond\cite{merinov2009proton,ma2026hydrogen}, thereby increasing the rotation barrier. In contrast, the lattice bending more easily forms the linear strong hydrogen bonds\cite{jing2020role,kreuer1996proton}, leading to a reduced transfer barrier. In Fig.~\ref{Fig10}(b), the correlation between the interoctahedral transfer barrier and the H-Ce distance reveals that the Ca-induced octahedral tilting elongates the H-Ce distance, thereby weakening the Coulomb repulsion and lowering the transfer barrier. Since rotation is the rate-limiting step in the BCO system, the influence of lattice contraction on the hydrogen bonds H...O$_f$ is dominant. The shortened hydrogen bonds increase the rotation barriers, which hinders proton diffusion in this system where rotation is the rate-limiting process.  This behavior is further supported by our lattice-dynamics analysis. From the $\Gamma$-point phonon calculations, we find that Ca doping increases the vibrational frequency of the stretching mode along the O-H bond direction from 3114.22 to 3392.06 cm$^{-1}$, while decreasing the frequency of the vibrational mode perpendicular to the O-H bond direction from 1253.86 to 1044.52 cm$^{-1}$. These changes suggest an enhanced tendency for proton transfer along the hydrogen-bond direction and a reduced tendency for proton rotation perpendicular to the O-H bond. In addition, Ca doping increases both the frequencies of O-Ce-O bending modes (295.53--378.25 cm$^{-1}$ to 298.37--383.5 cm$^{-1}$) and the Ba–CeO$_6$ stretching modes (34.51--289.17 cm$^{-1}$ to 49.48--296.37 cm$^{-1}$), reflecting a hardening of the lattice induced by the lattice-contraction effect. The resulting reduction in lattice flexibility further contributes to the increase in the proton rotation barrier\cite{muy2018tuning,krauskopf2018comparing}. However, for systems where transfer is the rate-limiting step, this effect becomes beneficial, as the reduced barrier of the rate-limiting transfer process and the smaller difference between transfer and rotation barriers can enhance proton conductivity\cite{bjorketun2005kinetic}. Therefore, A-site doping with small-radius cations such as Ca can be an effective strategy to improve proton diffusion in systems limited by the transfer process, consistent with prior reports\cite{yin2025breaking,dai2021tailoring,kang2025local}.

\subsection*{E. Chemical and Thermodynamic Stability of BCO and BCCO}
\begin{figure*}[htbp]
\centering
\includegraphics[scale=0.5]{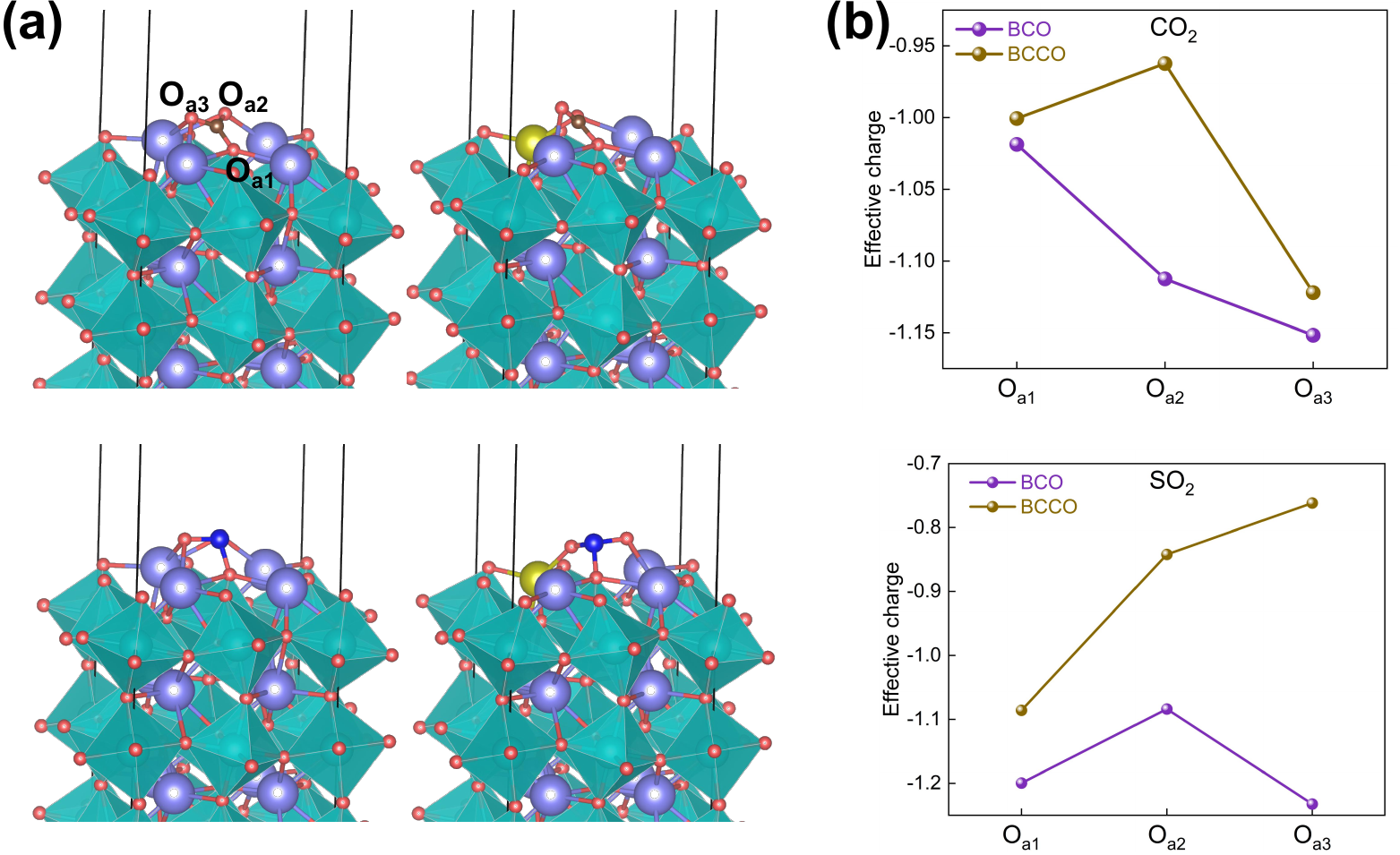}
\caption{(a) Top panels: carbonate structures formed by CO$_2$ adsorption on the BaO-terminated surfaces of BCO (left) and BCCO (right). Bottom panels: sulfate structures formed by SO$_2$ adsorption on the BaO-terminated surfaces of  BCO (left) and BCCO (right). The adsorption configuration of CO$_2$ is consistent with previous reports\cite{staykov2018interaction}, for example, an O–C–O angle of 123$^\circ$, C-O bond length of 1.35~\AA, Ba-O distance of 2.73~\AA, and a CO$_2^{3-}$ group tilted relative to the surface. (b) Valence states of the three oxygen atoms involved in the formation of the carbonates and sulfates shown in (a).}
\label{Fig11}
\end{figure*}

Experimentally, it has been observed that a small amount of Ca doping in BaCeO$_3$-based systems suppresses the formation of carbonate impurity phases under CO$_2$ atmospheres\cite{dudek2019ba0}. To elucidate the origin of the enhanced chemical stability of Ca-doped BCO in acidic environments, we calculated the adsorption energies of CO$_2$ and SO$_2$ molecules on the BaO-terminated surfaces of BCO and BCCO. The effect of Ca doping on the thermal stability of the systems is also predicted. The adsorption energy $E_{Ads}$ is defined as\cite{liu2020microscopic}:
\begin{equation}
E_{Ads} = E_{ads}-(E_{sur}+E_{gas})
\end{equation}
Here, $E_{ads}$ is the total energy of the carbonate or sulfate system formed by CO$_2$/SO$_2$ adsorption on the BaO-terminated surface\cite{staykov2018interaction} (see Fig.~\ref{Fig11}(a)),  $E_{sur}$ is the energy of the clean BaO-terminated slab, and  $E_{gas}$ is the energy of an isolated CO$_2$/SO$_2$ molecule, calculated by placing a single molecule in a cubic box with a side length of 15~\AA. The calculated adsorption energy of CO$_2$ on BCCO is -2.07 eV, which is higher than that on BCO (-2.26~eV). Using a larger 136-atom surface supercell, the CO$_2$ adsorption energies before and after Ca doping are -2.27 eV and -2.04 eV, respectively, excluding the influence of finite surface size.   Similarly, the adsorption energy of SO$_2$ on BCCO (-2.70 eV) is also higher than that on BCO (-2.89 eV). The calculated change of 0.19 eV is larger than those reported in other doping studies, which are typically below 0.1 eV\cite{peng2025enhancement,wang2026industrial,liu2020microscopic}. Therefore, Ca doping tends to reduce the chemical instability of the BCO system, which is the major drawback limiting its practical applications.

The increased adsorption energies of acidic gases can be attributed to the tendency of doped Ca atoms to segregate toward the surface, as confirmed by the slab energy calculations in Sec. A. Consequently, CO$_2$/SO$_2$ molecules are more likely to form CaCO$_3$ or calcium-based sulfates on the BCCO surface rather than BaCO$_3$ or barium-based sulfates on the BCO surface. As shown in Fig.~\ref{Fig11}(b), the calculated Bader charges of the three oxygen atoms involved in forming the surface carbonates and sulfates show that, in the calcium-based species, the oxygen atoms carry less negative effective charges, indicating weaker Ca-O bonds. As a result, the carbonates and sulfates formed on the BCCO surface are less stable than those on the BCO surface, leading to higher adsorption energies and reduced reactivity with acidic gases.

\begin{figure}[htbp]
\centering
\includegraphics[scale=0.3]{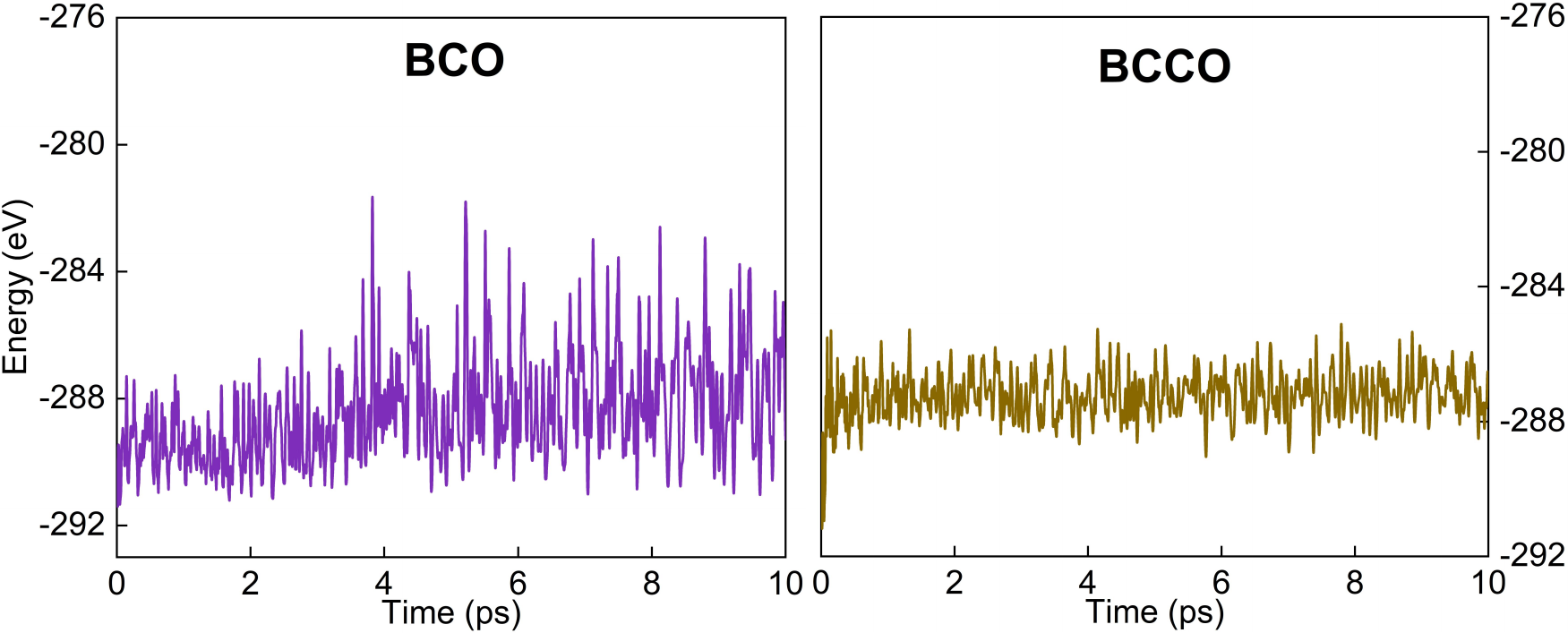}
\caption{Time evolution of the total energy of BCO (left) and BCCO (right) systems at 800 K from AIMD simulations.}
\label{Fig12}
\end{figure}

As shown in Fig.~\ref{Fig12}, 10 ps AIMD simulations were performed for BCO and BCCO at 800 K, a typical operating temperature for proton-conducting materials. The total energies fluctuate around equilibrium values without noticeable structural distortion, indicating the thermodynamic stability of both systems. More importantly, the amplitude of energy fluctuation in BCCO is significantly smaller than that in BCO, suggesting that Ca doping further enhances the thermal stability of BaCeO$_3$-based materials. This improvement mainly originates from the lattice contraction induced by Ca, which reduces the lattice flexibility, as evidenced by the increased frequencies of the O-Ce-O bending modes and Ba-CeO$_6$ stretching modes. Consequently, the vibrational amplitudes are suppressed, preventing temperature-induced strong structural distortions and bond breaking\cite{liu2020microscopic}, thereby improving the thermal stability.

\section{Summary}
To elucidate the experimentally observed enhancement of both proton conductivity and chemical stability in BaCeO$_3$-based electrolytes upon A-site Ca doping\cite{dudek2016some,dudek2019ba0,luo2023chemical,kothandan2024effect}, we comprehensively investigated the properties of Ca-doped BaCeO$_3$ electrolyte materials using DFT. Our results show that A-site Ca doping induces lattice contraction owing to the smaller ionic radius, while its higher electronegativity leads to weaker Ca-O ionic bonding compared with Ba-O bonding. By calculating the formation energies of oxygen vacancies in bulk and surface structures, we found that the weakened Ca-O bonds directly promote the formation of nearest-neighbor oxygen vacancies, while the lattice contraction effect indirectly facilitates the formation of next-nearest-neighbor vacancies. Due to surface reconstruction, Ca atoms located at the surface do not significantly enhance oxygen vacancy formation, which helps to mitigate grain-boundary resistance. Meanwhile, the weakened Ca-O bonding increase the basicity of the BCO system, leading to a more exothermic hydration reaction upon Ca doping. The reduced oxygen-vacancy formation energy and hydration enthalpy favor proton incorporation, thereby enhancing the proton conductivity. Further analysis of the proton migration barriers reveals that the lattice contraction induced by Ca doping increases the barrier of the rate-limiting rotation step, thereby hindering proton diffusion. However, this effect can be exploited in systems where the rate-limiting step is the proton transfer, promoting long-range proton diffusion. Moreover, the calculated adsorption energies of CO$_2$ and SO$_2$ show that the weakened Ca-O bonds suppress the formation of surface carbonates and sulfates, leading to higher adsorption energies and thus enhanced chemical stability of the BCO system under acidic environments. The AIMD simulations further demonstrate that the lattice contraction effect improves the thermal stability. Our results demonstrate the advantages of A-site Ca doping in Ba-based electrolytes, elucidate the mechanisms by which small-radius, high-electronegativity dopants enhance proton conductivity and chemical stability, and provide theoretical guidance for the design and optimization of high-performance proton-conducting electrolytes.

\noindent
\underbar{\bf Acknowledgements:}
This work was supported by Beijing Natural Science
Foundation (Nos.1252022 and 1242022), and National Natural Science Foundation of China (Nos. 12404463 and 12474218).

\noindent
\underbar{\bf Data availability:}
The data that support the findings of this article are openly available\cite{ma_2026_20725187}.

%%%%%%%%%%%%%%%%%%%%%%%%%%%%%%%%%%%%%%%%%%%%%%%%%%%%%%%%%%%%%%%%%%%%%%%%
%%%%%%%%%%%%%%%%%%%%%%%%%%%%%%%%%%%%%%%%%%%%%%%%%%%%%%%%%%%%%%%%%%%%%%%%
%Supplemental Material
%%%%%%%%%%%%%%%%%%%%%%%%%%%%%%%%%%%%%%%%%%%%%%%%%%%%%%%%%%%%%%%%%%%%%%%%
%%%%%%%%%%%%%%%%%%%%%%%%%%%%%%%%%%%%%%%%%%%%%%%%%%%%%%%%%%%%%%%%%%%%%%%%

\appendix
\section*{Appendix}

\setcounter{equation}{0}
\setcounter{figure}{0}
\renewcommand{\theequation}{A\arabic{equation}}
\renewcommand{\thefigure}{A\arabic{figure}}
\renewcommand{\thesubsection}{A\arabic{subsection}}

\subsection*{1. The PDOS of BCO and BCCO}
\begin{figure}[htbp]
\centering
\includegraphics[scale=0.37]{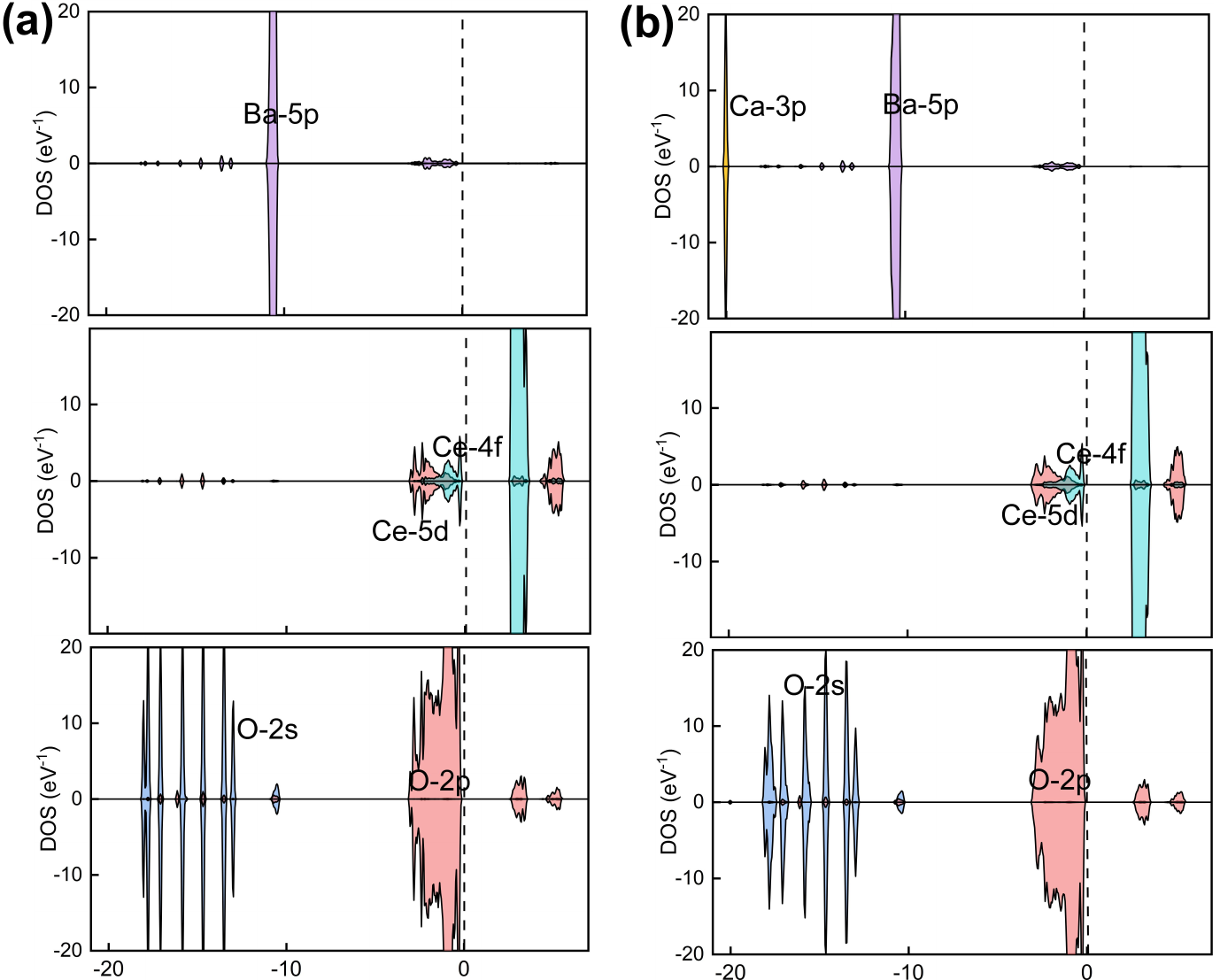}
\caption{The PDOS of (a) BCO and (b) BCCO}
\label{FigA1}
\end{figure}

From Fig.~\ref{FigA1}, it can be seen that in the BCO and BCCO systems, the Ce 4f and 5d orbitals exhibit strong hybridization with the O 2p orbitals near the valence band maximum (VBM), indicating the covalent nature of the Ce-O bonds. In contrast, the Ba 5p and O 2s states, as well as the Ca 3p and O 2s states, show only very weak orbital overlap at deeper energy levels. Therefore, the Ba-O and Ca-O bonds are predominantly ionic in character.

\subsection*{2.  The slab models with Ca located in the bulk and at the surface}
\begin{figure}[htbp]
\centering
\includegraphics[scale=0.5]{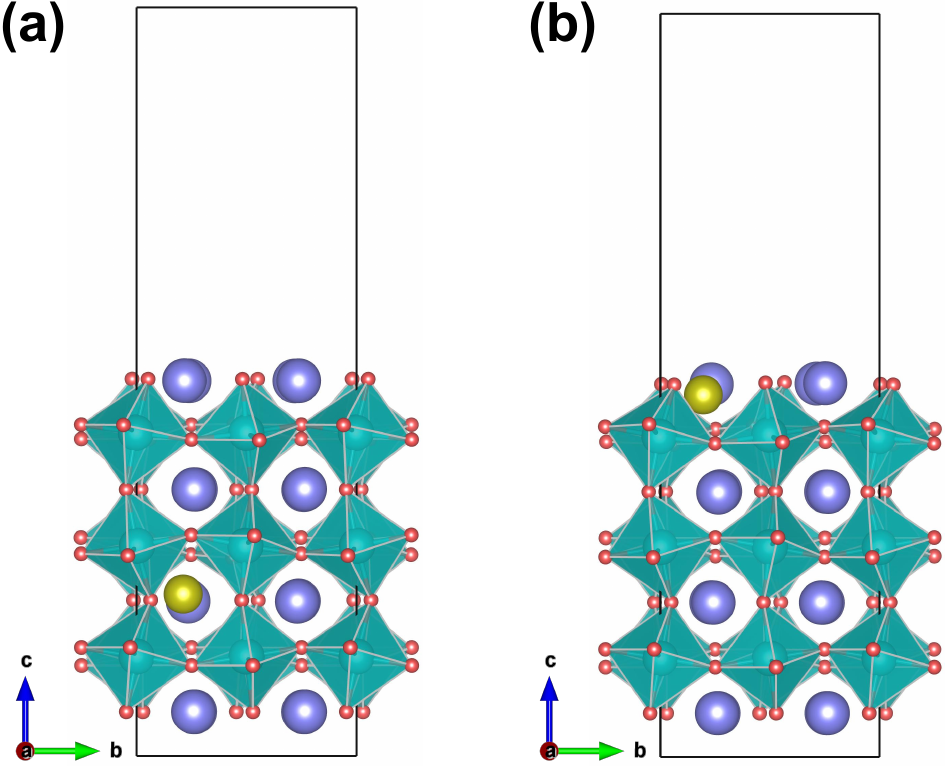}
\caption{BCCO slab models with Ca (a) located in the bulk and (b) segregated to the surface.}
\label{FigA2}
\end{figure}

As shown in Fig.~\ref{FigA2}, the total energy of the slab system with Ca doped in the bulk is -480.37 eV, whereas that with Ca segregated to the surface is -480.83 eV. Therefore, upon Ca doping into the BCO system, Ca tends to segregate toward the surface due to the weaker ionic bonding of Ca-O compared to Ba-O.

\subsection*{3. The structural reconstruction induced by Ca segregation to the surface}
\begin{figure}[htbp]
\centering
\includegraphics[scale=0.37]{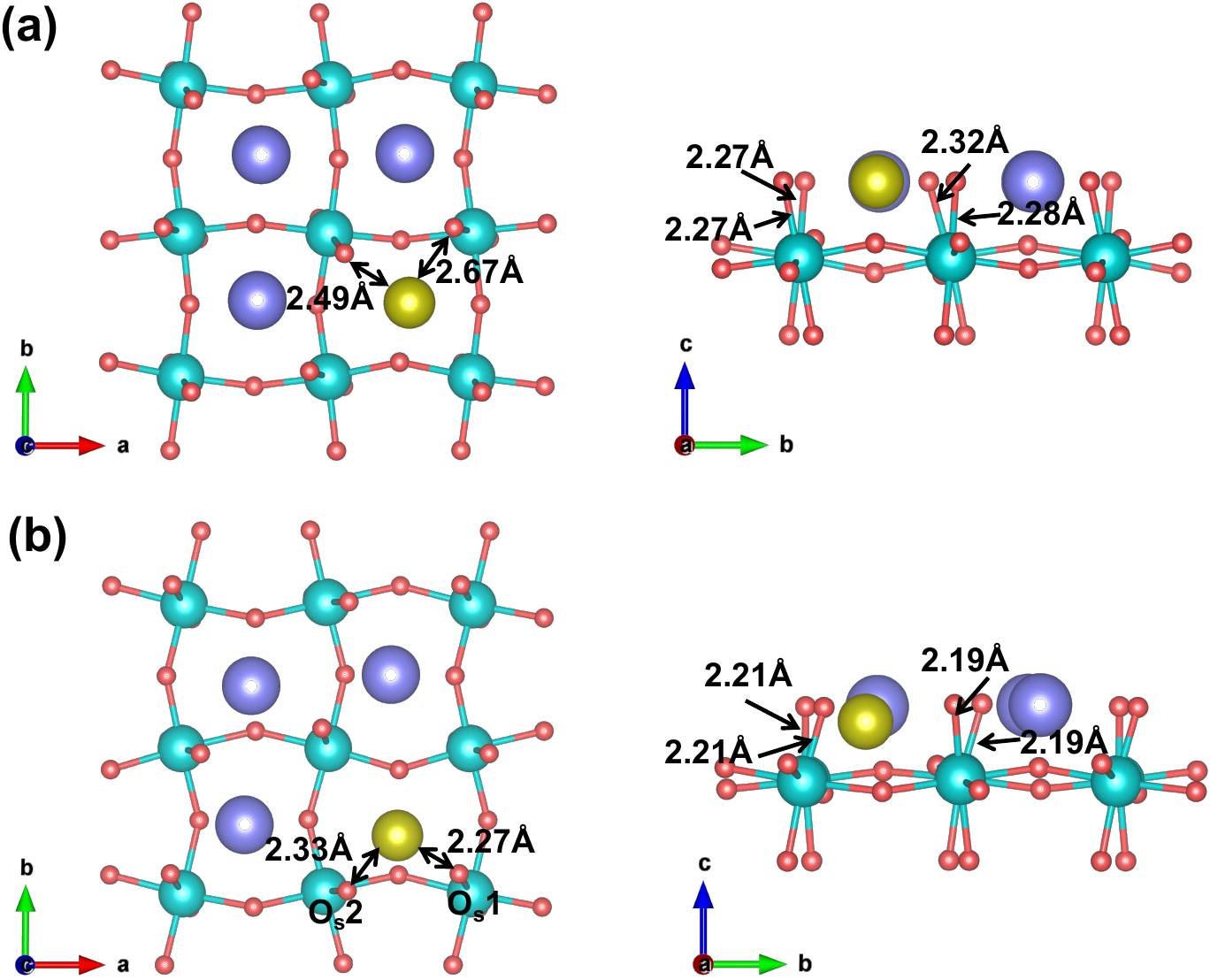}
\caption{(a) The (001) plane containing Ca in the bulk BCCO structure (left) and the side view of this (001) plane (right). (b) The (001) surface to which Ca segregates in the BCCO slab structure (left) and the side view of this (001) surface (right).}
\label{FigA3}
\end{figure}

Figure~\ref{FigA3} illustrates the lattice distortion induced by Ca segregation from the bulk to the surface. As shown in panel (b), the (001) surface exhibits more strongly tilted octahedra compared to the (001) plane in panel (a), which shortens the Ca–O$_s$1 and Ca–O$_s$2 distances and enhances the covalent character of the Ca-O bonds. Meanwhile, the Ce–O$_s$1, Ce–O$_s$2, Ce–O$_s$3, and Ce–O$_s$4 bond lengths are further reduced, strengthening the Ce-O bonds. These strengthened bonds explain the weaker promotion of oxygen vacancy formation when Ca segregates to the surface.

\subsection*{4. The charge density distributions of the unit cell for BCO and BCCO}
\begin{figure}[htbp]
\centering
\includegraphics[scale=0.3]{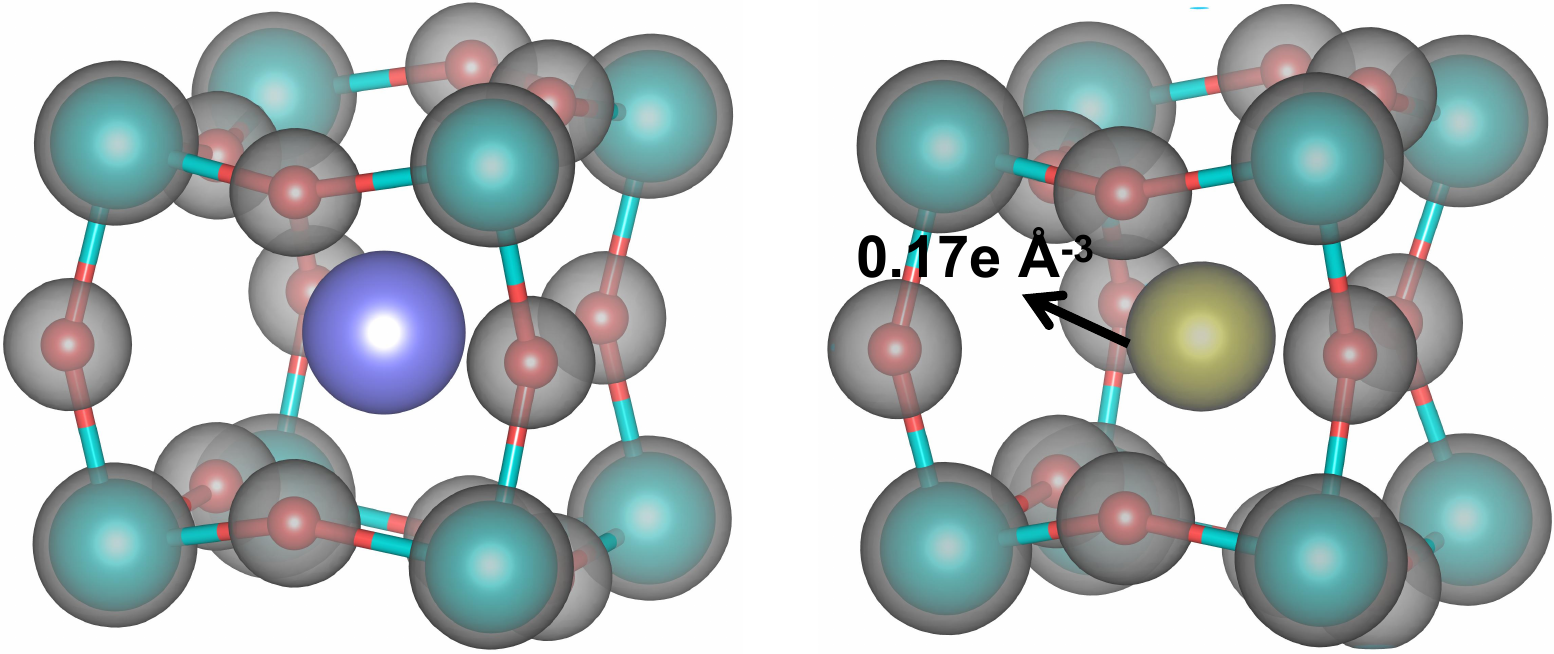}
\caption{The charge density distributions of the unit cell containing Ba in the BCO supercell (left) and the unit cell containing Ca in the BCCO supercell (right). Gray isosurfaces represent electron densities above 0.17e \AA$^{-3}$}
\label{FigA4}
\end{figure}

Figure~\ref{FigA4} illustrates the enhanced basicity of BCO upon Ca doping. At an isosurface level of 0.17e \AA$^{-3}$, a finite charge density appears around Ca in BCCO, whereas no comparable density is observed around Ba in BCO. This higher local electron density at the Ca sites renders them more electron rich, thereby favoring proton uptake and promoting the hydration reaction.

\bibliography{ref}

\end{document}